# Main Manuscript for

Multistable Topological Mechanical Metamaterials

Haning Xiu[1], Harry Liu[2], Andrea Poli[3], Guangchao Wan[4], Ellen M. Arruda[3,5,6], Xiaoming Mao[2], Zi Chen[1,*]

[1]Brigham and Women's Hospital, Harvard Medical School, Boston, MA 02115, USA

[2]Department of Physics, University of Michigan, Ann Arbor, MI 48109, USA

[3]Department of Mechanical Engineering, University of Michigan, Ann Arbor, MI 48109, USA

[4]Department of mechanical and aerospace engineering, Syracuse University, Syracuse, NY 13210, USA

[5]Department of Biomedical Engineering, University of Michigan, Ann Arbor, MI 48109, USA

[6]Macromolecular Science and Engineering, University of Michigan, Ann Arbor, MI 48109, USA

* Zi Chen, 75 Francis St, MA 02115, +1(609)356-9441, **Email:** zchen33@bwh.harvard.edu

**Author Contributions:** Haning Xiu contributed to build multistable Maxwell lattice models and 3D printing, numerical simulation, experimental implementation, and manuscript preparation. Harry Liu contributed to numerical simulation, multi-material Maxwell lattice design and experiments, and manuscript preparation. Andrea Poli contributed to multi-material Maxwell lattice design and experiments. Guangchao Wan contributed to build 3D bistable unit cell modelling of Maxwell lattice. Xiaoming Mao and Zi Chen conceptualised the project. Xiaoming Mao, Ellen Arruda, and Zi Chen contributed to developing the prototype, modelling, data analysis, and manuscript editing.

**Competing Interest Statement:** The authors declare no competing financial interests.

**Classification:** 1) Applied Physical Sciences; 2) Engineering—Mechanical engineering

**Keywords:** Maxwell lattice, Bistability, Stiffness tuning, Topological transformation, 3D printing.

**This PDF file includes:**

> Main Text
> Figures 1 to 6

## Abstract

Concepts from quantum topological states of matter have been extensively utilized in the past decade in creating mechanical metamaterials with topologically protected features, such as one-way edge states and topologically polarized elasticity. Maxwell lattices represent a class of topological mechanical metamaterials that exhibit distinct robust mechanical properties at edges/interfaces when they are topologically polarized. Realizing topological phase transitions in these materials would enable on-and-off switching of these edge states, opening unprecedented opportunities to program mechanical response and wave propagation. However, such transitions are extremely challenging to experimentally control in Maxwell topological metamaterials due to mechanical and geometric constraints. Here we create a Maxwell lattice with bistable units to implement synchronized transitions between topological states and demonstrate dramatically different stiffnesses as the lattice transforms between topological phases both theoretically and experimentally. By combining multistability with topological phase transitions, for the first time, this metamaterial not only exhibits topologically protected mechanical properties that swiftly and reversibly change, but also offers a rich design space for innovating mechanical computing architectures and reprogrammable neuromorphic metamaterials. Moreover, we design and fabricate a topological Maxwell lattice using multi-material 3D printing and demonstrate the potential for miniaturization via additive manufacturing. These design principles are applicable to transformable topological metamaterials for a variety of tasks such as switchable energy absorption, impact mitigation, wave tailoring, neuromorphic metamaterials, and controlled morphing systems.

**Significance Statement**

Mechanical metamaterials exhibit exotic properties governed by structures rather than constituents. They provide a unique platform to recapitulate atomistic structural arrangements at macroscale and thus achieving advanced functionalities hitherto impossible. Maxwell lattices, a class of topological mechanical metamaterials, offer tunable mechanical properties and topological transitions with predicted applications in energy absorption, impact mitigation, and waveguiding. However, such transitions are extremely challenging to control and synchronize between units. Here we create a multistable topological Maxwell lattice combining mechanical multistability with topological phase transitions and demonstrate swift, reversible topological transitions, whereby surface mechanical properties drastically change. This design will unleash the potential of the applications of mechanical metamaterials in stiffness tuning, impact mitigation, mechanological and neuromorphic computation, information processing, and flexible robotics.

**Main Text**

**Introduction**

Synthetic mechanical metamaterials have been demonstrated to exhibit exotic physical and mechanical properties such as negative Poisson's ratio[1], negative compressibility[2], and programmable nonlinearity[3, 4]. One group of mechanical metamaterials, Maxwell lattices, has been extensively studied in recent years due to their proximity to mechanical instability and interesting topological edge floppy modes and states of self-stress[5–8], leading to a wide collection of fascinating phenomena and applications including focused stress and fracture protection, phonon waveguide, phonon diode, and tunable stability[3, 9–15]. The distorted kagome lattice, one of the typical Maxwell lattices, has been the focus of extensive research due to its rich mechanical properties in various topological states[16–21]. By tuning the geometry of unit cells, a distorted kagome lattice can exhibit peculiar mechanical behaviour protected by the topology of the compatibility matrix in momentum space. Recent research[5] showed the feasibility of changing the topological phase of a kagome lattice by varying a single geometric parameter, i.e., the angle between the two triangles in the unit cells; such transformation is done using a homogeneous soft strain that is intrinsic to the lattice, called the Guest-Hutchinson Mode[22]. The topological transition induced by this soft strain leads to dramatic changes in the surface stiffness of the lattice, desirable for many applications from impact mitigation to non-reciprocal wave propagation[23]. However, experimental realization of this soft strain transformation has been challenging, due to the proximity of these lattices to mechanical instability, which leads to inhomogeneities in deformation as well as nonlinear buckling of the lattice geometry during the transformation. This challenge makes it difficult for Maxwell lattices to become usefuly multi-functional materials and potential innovative reprogrammable mechanological metamaterials (ReMM)[24–26].

To address this, here we introduce bistability, for the first time, to topological Maxwell lattices by adding a spring to each unit cell. This novel design achieves swift and homogeneous transformation between topologically polarized[27] and non-polarized phases, characterized by dramatically different mechanical properties, and meanwhile it generates programmable mechanical stiffnesses including effective elastic stiffness and shear stiffness that do not exist in conventional Maxwell lattices. In the design we present here, the added spring connects the centers of two bonds in each unit cell, thereby creating two stable states that lie symmetrically on either side of the critical state that separates the topologically polarized and nonpolarized phases (Fig 1). Both theoretically and experimentally, with good agreement, we show that the surface stiffness in one state is significantly different from the other. We then characterize how the spring constant of the added springs controls the stiffness of the soft surface. We further demonstrate experimentally that the bistability of these units greatly improves the controllability of the lattice during the transformation from one state to the other, realizing swift, drastic, and reversible change of surface stiffness.

Beyond these distinct characteristics, such as maintaining mechanical properties of conventional Maxwell lattices and generating mechanical stiffnesses, these multistable Maxwell lattices open a suite of new opportunities for combining topological mechanics with nonlinearity. The multistable lattice we designed shows intriguing spatial patterns that feature drastically different structures separated by an interface; each side adopts one of the two stable states whereby the interface depends on various factors from the width of the strip to the unit cell design, creating a new set of controllable spatially patterned materials. The proposed multi-stable Maxwell lattice can be further used to design and innovate ReMM. It will establish a new foundation that will facilitate the development of mechanical computing systems, neuromorphic metamaterials, controlled morphing systems, flexible robotics, bioinspired control, and distributed intelligence.

**Results**

*Topological polarization and multistability of Maxwell lattices*
Mechanical stability of a frame to linear order can be characterized using zero modes (ZMs) and states of self-stress (SSSs). A ZM is a set of displacements of the lattice sites that cause no change of bond length, and an SSS is a set of tension/compression on bonds that cause no net force on lattice sites. Maxwell proposed, and Calladine later modified, an index that accounts for the difference between the degrees of freedom of a network and the number of constraints, which is equal to the difference between the number of ZMs and SSSs[6, 27, 28]. In the special case when the number of degrees of freedom exactly equals to the constraints in the bulk, the mechanical network is at the verge of instability, and such a network is called a Maxwell network. For lattices with point-like sites and central-force bonds ("ball-and-spring" structures) this criterion takes the form that the average coordination number of the network is $\langle z \rangle = 2d$, where $d$ is the dimension of the network, leading to $\langle z \rangle = 4$ in two dimensions. One of the typical lattices that satisfies the Maxwell criterion is the kagome lattice and its various distorted versions, which contains 3 sites and 6 bonds in its unit cell, satisfying $\langle z \rangle = 4$ (each bond connects two sites). In Fig. 1a, one such unit cell is shown as a combination of triangles A and B, and the specific side lengths of triangles A and B we use in this paper are (0.4, 0.8, 1), (0.5, 0.7, 1). In the limit that the bonds of the lattices are rigid, only one internal degree of freedom, the relative rotation between the two triangles (the twisting angle), $\alpha$, remains for this unit cell.

Mechanics of Maxwell lattices are governed by a "topological polarization" $R_T$ introduced by Kane and Lubensky[6]. This topological polarization is determined by the geometry of the unit cell and can be computed via winding numbers of the compatibility matrix of the lattice in the momentum space. Boundaries and interfaces where $R_T$ points towards gain topologically protected, exponentially localized ZMs, and by contrast, boundaries and interfaces where $R_T$ points away gain SSSs. Such topological polarization leads to remarkable mechanical effects owing to asymmetric surface stiffness and stress focusing[10, 15]. In Ref. [5], it is shown that changing the geometry of the unit cell via the Guest-Hutchinson mode (which alters the angle between triangles A and B homogeneously across the lattice) changes the topological polarization, thus switching the mechanical properties of the boundaries and interfaces.

In particular, this distorted kagome lattice experiences such topological transitions at three critical angles, $\alpha_{a_2}$, $\alpha_{a_1}$, and $\alpha_{a_2-a_1}$ (Fig. 1b, where subscripts of $\alpha$ are the lattice directions where bonds align into straight lines at the transition). When $\alpha < \alpha_{a_2}$ or $\alpha > \alpha_{a_2-a_1}$, the lattice is non-polarized ($R_T = 0$, where all boundaries are soft), and as $\alpha$ changes between $\alpha_{a_2}$ and $\alpha_{a_2-a_1}$, the lattice exhibits a topological polarization. There are two specific polarized phases where $R_T = a_2$ and $R_T = a_2 - a_1$ separated by the critical angle $\alpha_{a_1}$.

To achieve multistability of the lattice and convenient operation of topological polarization transformation, bistable units (springs with low stiffness shown by black lines in Fig. 1a and c) are added to connect the centres of two bonds in each unit cell. Fig. 1c indicates the two stable equilibria of a unit cell: one is located in the polarized phase, and the other in the non-polarized phase (3D printed bistable unit cell is shown in Fig. 1d). The added bistability of each unit facilitates the lattice transition between the two equilibria via the Guest-Hutchinsin modes. As the rest length of the springs, which is lower than the half length of $(a_b + b_r)$, changes, the angle of the homogeneous zero-energy configuration of the lattice, $\alpha$, is tuned. When the bistable lattice goes through a transition to the polarized phases, there is a dramatic increase of the edge stiffness at the edge opposite to the polarization direction (Fig. 1e), preserving the topological effect observed in lattices without bistable units[5]. Moreover, these bistable units dramatically decrease the number of ZMs of the lattice with n × m unit cells from $2n + 2m - 3$ to 2 (rigid rotations of the dangling triangles in the top-left and bottom-right unit cells). Although this change effectively rigidifies the lattice, we show below that because these springs are much softer than the triangles, the lattice is still close to the Maxwell point and the topological properties are well-preserved[29]. Theoretical analysis, numerical computations, and experimental validation of this bistable design will be discussed in the following section to characterize how mechanical properties of the lattice are affected by the springs.

### *Surface mechanical properties of multistable Maxwell lattices*

In this section we discuss the effects of the springs, which are added to induce bistability, on the surface mechanical properties of the lattice and on the generated mechanical stiffnesses for the lattice. We show that, interestingly, the stiffness of the soft edge increases linearly with the stiffness of the spring, whereas the stiffness of the rigid edge is insensitive to the spring constant. Additionally, the springs generate mechanical stiffness when transforming the lattice from one state to another.

A Maxwell lattice of n × m unit cells is created to quantify the surface stiffness of the lattice during the topological polarization transitions. To investigate how the bistable units affect the mechanics of the lattice, linearized surface stiffnesses are calculated and compared with the results of the same lattice without bistable units (Fig. 3b). First, we start from the nonlinear elastic energy of the lattice by fixing the rest length of the spring $l_s$ (for a given angle $\alpha$). It is written as

$$T = \sum_i \sum_{j=1}^{6} \frac{k_1}{2} (l_{ij} - l_{ij0})^2 + \sum_k \frac{k_2}{2} (l_k - l_s)^2, \qquad (1)$$

where $i$ and $k$ label the unit cells and $j$ labels the 6 bonds (edges of triangles) in each unit cell, $l_{ij0}$ and $l_{ij}$ are the rest and current lengths of each bond, respectively, $l_k$ represents the length of the spring in its current state. The stiffnesses of the bonds and the added springs are taken to be $k_1$ and $k_2$ respectively. In the limit of small strain on the bonds and springs the elastic energy of the lattice (energy of each bond and spring) can be linearized. More details about the elastic energy and its linearization in a unit cell can be found in the Supplementary Note. To perform the test, three out of four boundaries of the lattice are fixed, and the remaining boundary is free to be given a small displacement at the center (Fig. 2a-b), the direction of which is perpendicular to the bottom edge ($a_1$ vector direction). The linear conjugate gradient method is used to solve the minimal energy of the lattice and the surface stiffness is calculated as the ratio between force and displacement for the node being pressed.

The springs not only introduce bistability, but also generate mechanical stiffness for the lattice, such as effective elastic stiffness and effective shear stiffness of a 2D lattice; here the effective elastic stiffness, $k_{el}$, is defined as the ratio between the force and the displacement in the direction perpendicular to the primitive vector of $a_1$, when the lattice undergoes uniform twisting, and the effective shear stiffness, $k_{sh}$, is defined as the ratio between the force and the displacement in the direction parallel to $a_1$:

$$k_{el} = k_2 \frac{b_r^2 \sin^2(\alpha+\psi_{ar})}{8l_s^2 \cos^2(\pi-\psi_{bb})|\cos\phi|}, \quad (2)$$

$$k_{sh} = k_2 \frac{b_r^2 \sin^2(\alpha+\psi_{ar})}{8l_s^2 \sin^2(\pi-\psi_{bb})|\sin\phi|}, \quad (3)$$

where $b_r$ is a side length of triangle A, $\psi_{ar}$ and $\psi_{bb}$ are inner angles of the triangles, and $\phi$ is the angle between the vector of $a_1$ and the side of $c_b$. More information on the mechanical stiffness is shown in the Supplementary Note. Here, the elastic and shear stiffnesses have linear relationships to the spring constant $k_2$ and are also dependent on the initial twisting angle $\alpha$, shown in Fig. 2g. When $k_2 = 0$, both $k_{el}$ and $k_{sh}$ are zero and the lattice transformation between topological states is a purely soft one as discussed in Ref. [5].

The surface stiffness of the lattice depends on the rest length and spring constant of the added springs, as well as the configuration of the lattice. We first study the effect of the spring's rest length on the surface stiffness. By varying the rest length of the springs between 80%-99% of $(a_b + b_r)/2$, we tune the two equilibria between far from the topological transition to close to the transition (Fig. 2c). We use the linearized method described above to measure the edge stiffness of a 30 × 30 multistable lattice at different rest lengths and different $k_2$ values (Fig. 2d-e), starting from the topologically polarized equilibrium state. Our results show that the stiffnesses of the soft and hard edges are not sensitive to the rest length of the springs, as long as the lattice is placed at its equilibrium in the polarized phase.

We then study the effect of changing spring constant $k_2$ on the edge stiffness (Fig. 2f). We found that the stiffnesses for both hard and soft edges monotonically increase as $k_2$ increases, but the soft edge is more sensitive to $k_2$. The magnitude of $k_2$ ranges from $10^{-4}$ to 100 and the stiffness of bonds stays at $k_1 = 1$. Although the edge stiffnesses for different rest lengths show moderate variations, the stiffness for the hard edge increases very slowly when $k_2$ increases up to $10^{-1}$, and then it steeply increases as $k_2$ continues to increase. The stiffness of the soft edge shows the opposite trend—it increases exponentially when $k_2$ is from $10^{-4}$ to slightly less than 1 and then increases less sharply afterwards. Because of the different increase of stiffnesses between hard and soft edges, Fig. 2f also represents the stiffness convergence of both edges at large $k_2$ values (up to 100) for each case. This trend can be understood via the nature of the topological edge modes. In the Maxwell limit ($k_2 = 0$), the edge modes are purely rotations of the triangles. At the soft edge, these are the modes that accommodate the local deformations. As a result, the edge stiffnesses are extremely sensitive to the spring constant $k_2$, because the topological edge modes distort these springs. At the hard edge, in contrast, there are no topological edge modes, and the local deformation is composed of bulk modes, which distorts both the bonds and the springs in the kagome lattice. When the bonds $k_1$ are much stiffer than the springs, the hard-edge stiffness is not sensitive to $k_2$. As $k_2$ increases to close to or even greater than $k_1$, the added springs become comparable to the bonds in the unit cells, which will take the lattice away from Maxwell points, and lead to the disappearance of topological polarization of the lattice and the lattice will transition to a conventional stable elastic medium.

We further study the edge stiffness of a Maxwell lattice as it deforms along its Guest mode (labelled by the twisting angle $\alpha$), which reconfigures the lattice between topological phases (Fig. 3). The general shape of these curves is similar to the case of lattices without bistable units (Fig. 3b). However, the addition of the lattice is only in force balance at the two equilibria (red and black dots in Fig 3), due to these bistable units. In all other configurations, the lattice is held at the boundary, and then relaxed to adopt an inhomogeneous and stressed reference state (via a minimization of the nonlinear elastic energy). We then expand the elastic energy around these inhomogeneous stressed states and use the linearized theory to calculate the edge stiffness. The analysis of the boundary stiffness shows that the topological polarization of Maxwell lattices is well-preserved upon the incorporation of the springs. Furthermore, both the surface stiffness and the lattice elasticity can be programmed by tuning the properties of these added springs.

### *Reconfigurable interface between topologically distinct states of a multistable Maxwell lattice*

The introduction of multistability to a Maxwell lattice offers new opportunities for studying intriguing structural features from topological interfaces to defects, as well as their reconfigurability. As an important first step in understanding heterogeneous states in multistable Maxwell lattices, here we study the static interface between topologically polarized and unpolarized phases in a multistable lattice.

For a homogenous multistable Maxwell lattice, as a shear strain is applied from one boundary of the lattice, a portion of the lattice undergoes a topological transition from one topological state to another. However, due to the different primitive vectors between the two lattice domains in different topological states (Fig. 4i), the interface bears geometric frustration, and interesting questions arise in terms of the equilibrium configurations of the lattice. To understand the profile of the interface in the static limit, a numerical study using energy minimization is conducted. We set up the initial configuration such that two lattice domains in distinct topological phases, namely, one unpolarized ($R_T = 0$) and one polarized ($R_T = a_2 - a_1$), are connected. Depending on which domain is placed on the top, we have either a SSS interface or a ZM interface[6]. As mentioned above, because the lattice vectors of the two domains along the interface are in general different, in order to connect them, we elongated some bonds in the initial condition, and then let the lattice evolve according to (nonlinear) energy minimization to reach the equilibrium configuration. As a result, interesting interface configurations show up, as shown in Fig. 4.

The lattice-vector mismatch between the two domains causes interesting profiles where the interface curves and widens in all cases. This curvature shares similarities with curved thin elastic bodies in continuum elasticity due to frustration[30–32]. Another key feature is that the exact interface profile depends on the initial configuration, i.e., how the two domains are connected initially, because the energy minimization could lead to multiple local minima given the complexity of the energy landscape. The initial configuration to create interface profiles in Fig. 4 is described in the supplementary materials.

The study of the interface profile reveals a geometric frustration when this multistable lattice transforms from one topological state to another via an interface. When the whole lattice transforms uniformly, there is no geometric frustration, but as we described above, when this transition occurs across the lattice with an interface, the lattice vector mismatch causes frustration and internal stress, giving rise to interesting profiles. While such a transition is often dynamic (i.e., as the lattice is triggered by an impact from one end and the transformation propagates through the lattice), we have chosen to focus on the statics in this study; the dynamics of this interface profile will require further study to achieve a deeper understanding of how this geometric frustration interplays with waves of stress.

### *Experiments on multistable hinged Maxwell lattices*

A prototype of a 3D-printed lattice is used for experimental demonstration of the theoretical findings in the previous sections. The lattice consisting of 6-unit cells in both length and width are assembled with bistable units, where the added springs are incorporated between two triangles from adjacent unit cells (one unit cell with spring embedded is shown Fig. 1d). Fig. 5a shows the combined lattice transitioning from the non-polarized domain to the polarized domain via soft twisting manually[5]. This process experimentally demonstrates the ease of manipulation of topological transformation, where the lower left of the lattice is held by one hand and the upper right corner can be pulled diagonally using the other hand. Compared with the similar manipulation in the previous study[5], where the Guest mode has to be realized by a uniform dilation which is difficult to control, the multistable lattice efficiently improves the twisting manipulation and speeds up the transition process. Supplementary video 1 shows the entire transformation process of the lattice.

The experimental setup for measuring the surface stiffness of the lattice is shown in Fig. 5b and the corresponding force/stiffness versus displacement results for two opposite edges are given in Fig. 5c and d. Tips on three of four edges are bonded to the ground using hook-and-loop self-adhesive tapes (also adding superglue to prevent edges slipping) and the last edge is a free boundary for applying mechanical loads for testing. Fig. 5b demonstrates the large pushing force needed (43.7 N) when a displacement (here 20 mm) is given at a tip of the hard edge, as well as the small deformations as a result, while a more distinct shape change is observed when it is pushed at the tip from the soft edge, and the corresponding force (0.86 N for 19 mm) is roughly 2% of the one at the hard edge. The edge stiffness (the ratio of force to displacement) is calculated and presented in Fig. 5c. The mechanical test is implemented within the regime where the hinges start to bear mechanical forces. More details of the differences between theoretical and experimental stiffness are explained in the Supplementary Note. Fig. 5d illustrates that there are roughly two orders of magnitude difference between the stiffness of the hard and soft edges (when a pushing load is given). The experimental results are further proofs that the polarized multistable lattice maintains the property where the surface stiffness of the hard edge sharply increases and becomes much higher than that of the soft edge.

Through a set of 3D-printed Maxwell lattices, it is verified that the transformation of the lattice can be done through merely soft twisting of the network and the final angles of the unit cells are guaranteed to be nearly homogenous as they are controlled by the energy minimum of the springs. Furthermore, since the force-displacement curves are measured on both sides of the lattice as it falls in the topologically polarized state, the difference in the edge stiffness on those two sides due to the topological polarization is observed, as well as the hardening due to the addition of the springs.

***Experiments on multi-material Maxwell lattice and hingeless bistable unit***
To further demonstrate a significant contrast of the edge stiffnesses between two boundaries of the lattice, a multi-material additive manufacturing technique is utilized. In this design, VeroWhite, a stiff material, has been used to construct a solid triangle body for the unit cells to maintain the unit cell structure, and Agilus30, a softer material, has been used as the hinges connecting the triangles in the lattice. The hinges are stiff in response to a tensile force along the hinge axis because the Agilus 30 hinge is constrained via its attachment to the stiff Vero-White triangles. The hinges are compliant in bending because the soft Agilus 30 is free to bend. During the force vs. displacement measurements of the multi-material lattice, the boundary of interest is probed by a force sensor while the other 3 boundaries are fixed in place. The measurement shows a promising contrast between the edge stiffnesses of the rigid and floppy edges with an average initial stiffness ratio of ~16. The experimental set up and the measured force-displacement curves are shown in Fig. 6.

This experiment demonstrates that the incorporation of multi-material 3D printing nicely preserves the topological polarization, which was originally defined in the "pin-joint" limit with no bending stiffness. Homogeneous (one-material) 3D printing of Maxwell lattices led to the design of interesting metamaterials for contrasting surface stiffnesses[33–35]. However, the ratio of the surface stiffness at the hard and soft edges are limited to ~<5 in these designs, due to significant bending stiffness at the hinges. Multi-material 3D printing presented here offers a creative way to resolve this issue, using soft materials at the hinges, as discussed above. Furthermore, the softness of these hinges may also enable reconfiguration of the lattice between topological phases and the incorporation of the bistable mechanism, and these will be topics for future studies.

With a desire to achieve bistability in the Maxwell lattices through additive manufacturing, a new design of a hingeless bistable unit (Fig. 6c) is proposed to be incorporated to connect the red and blue triangles in each unit cell. A force displacement test was performed to show the transition between the two stable states of the unit in Fig. 6d. The fact that the force curve drops down to zero and then rises back up as the unit is pushed in a single direction indicates the existence of the second stable state.

However, the integration between the bistable units and the topologically polarized kagome lattice has not been achieved yet due to both size limitation and material properties. A lattice with small unit cells such as the one shown in Fig. 6a cannot accommodate the bistable units as they would be under the printing resolution of the 3D printer, while enlarging the unit cells to incorporate the bistable units would lead to an overall size of the lattice beyond the limitation of the printing platform.

A further engineering design needs to be considered to succeed in the integration of bistability and topological kagome lattices through additive manufacturing methods.

**Discussion**

In summary, a multi-stable topological mechanical lattice is first-ever created by attaching a spring between two specific edges of the unit cell of a topological kagome lattice. We experimentally and computationally show that the new design of the multistable Maxwell lattice can achieve a controllable and swift transformation between different topological phases with dramatically different mechanical behaviors and enhanced load-bearing capacity (Fig. 2f). We also studied interfaces between domains of lattices in different topological phases, revealing rich patterns. The asymmetric dynamics of this interface induced by the topological polarization will be interesting topics for future studies, with promising applications in impact control. In addition, we explored additive manufacturing methods to produce multistable lattices, demonstrating that multi-material 3D printing can help achieve edge stiffness ratios much greater than in previous studies, as well as proposing a bistable hingeless unit which can be incorporated into these lattices in future studies.

Multi-stable Maxwell lattices have broad potential applications according to different working circumstances. The edge of the polarized Maxwell lattices with a higher stiffness (hard edge) can be used for ballistic protection and vehicle components acting as load-bearing elements, while the soft edges of the lattices can be used as energy absorption layers and buffers. As the lattice reprograms to the non-polarized phase, both edges become soft, and this effect can be utilized to absorb and store external impact energy. The introduced bistability enables a feasible approach for improving the operation of programmable topological mechanical metamaterials. While in this specific example, bistablity in each unit cell is achieved through an addition of a spring, there are other methods by which bistability or multistability can be introduced including but not limited to strain engineering[36–39] and mechanical buckling[40], opening important areas for future study. Additionally, the multistable mechanical metamaterials, with novel mechanical properties and deformation modes, provide a new approach to innovate mechanical computing architectures (both non-volatile and volatile systems). Hence, our study can inspire neuromorphic metamaterials with unprecedented reprogrammability and controllability, for instance, ReMM that can perform advanced logical functions thanks to the multi-stability and superb reconfigurability of the lattice.

**Materials and Methods**

The primary methods used in this paper were theoretical analysis combined with experimental validation. Computer-aided programming and numerical computations were implemented via MATLAB. The prototype consisting of 6 × 6-unit cells was 3D printed using PLA materials by a commercial 3D printer (Ender 3 Pro) and assembled with springs (Uxewll 304 stainless steel) to generate a multistable lattice. A surface stiffness test was implemented using a digital force gauge by pushing and pulling tips at the soft and hard edges.

The experiment on the multimaterial-3D-printed lattice was performed on a platform that employs a constant-speed actuator instrumented with a 1N force sensor (FORT 100). Three boundaries of the sample were fixed on the platform by clamps and tapes, and the force probe indented a selected tip at the open boundary by up to ~3 mm.

The hingeless bistable unit was 3D printed with a Nylon material (Ultracur3D® ST 45) on a commercial 3D printer (Origin MDK4). The mechanical characterization of the bistable hinge was performed on a dynamic mechanical analyzer (TA RSA3) using a custom fixture to apply pressure to the outer edges of the unit while providing clearance for the spring element.

**Acknowledgments**

This work was supported by the Office of Naval Research (MURI N00014-20-1-2479).

**Figures**

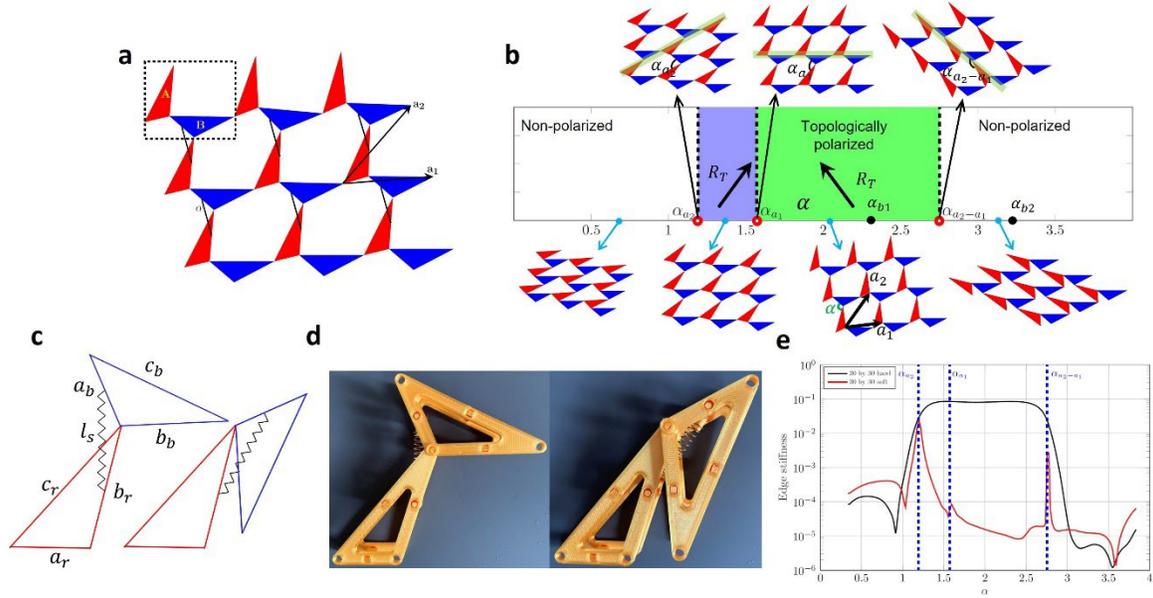

**Figure 1.** Schematic of the multistable Maxwell lattice and its topological polarization. (**a**) Schematic of a section of a Maxwell lattice having bistable units with 3 × 3-unit cells. Each unit cell (dashed box) consists of triangles A and B that are hinge-connected, and bistable units are indicated by black lines between the triangles. The primitive vectors are $a_1$ and $a_2$. (**b**) Topological transitions of a distorted kagome lattice. Three angles, $\alpha_{a_2}$, $\alpha_{a_1}$, and $\alpha_{a_2-a_1}$ correspond to cases where the sides of the triangles A and B align (light green stripes) and represent the critical angles that separate distinct topologically polarized phases. (**c**) Two stable equilibria with notations of the triangles of a bistable unit with a spring. $a_r$, $b_r$, and $c_r$ are side lengths of triangles A, while $a_b$, $b_b$, and $c_b$ represent side lengths of triangles B. $l_s$ is the length of the undeformed spring. (**d**) 3D printed unit cell with a spring showing bistable equilibria. (**e**) Numerical surface stiffness versus the twisting angle, $\alpha$, of a 30 × 30 distorted kagome lattice [as shown in (**a**)] with bistable units (spring constant $k_2 = 0.0001$ and bond stiffness $k_1 = 1$). The stiffness of the hard edge increases significantly as $\alpha$ falls within $\alpha_{a_2}$ and $\alpha_{a_2-a_1}$, and floppy modes move the opposite edge.

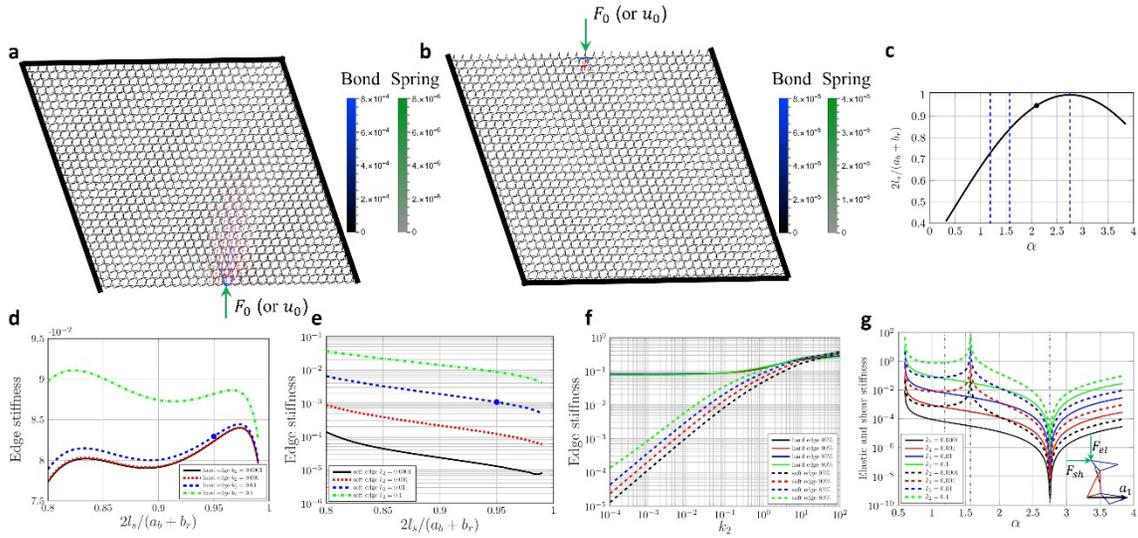

**Figure 2.** Parametric study of surface stiffness. Deformed lattice of 30 by 30-unit cells given displacement at the (**a**) hard (bottom of the lattice) or (**b**) soft (top of the lattice) boundary for calculating edge stiffness. The displacement field of the nodes are represented by red arrows with longer arrows corresponding to larger displacements. The tension/compression on the bonds and springs are shown in blue and green respectively with higher magnitudes corresponding to more visible colors and thicker lines. The springs are at their rest length, which is 95% of $(a_r + b_b)/2$, and $k_2 = 0.001$. (**c**) Rest length of springs $(2l_s/(a_b + b_r))$ required to make the homogeneous lattice in equilibrium at different $\alpha$ angles. The black dot is the configuration where the rest length is 95% of $(a_r + b_b)/2$, and this configuration is also used in Fig. 3. Surface stiffness of (**d**) hard and (**e**) soft edges against the rest length of the bistable unit for different spring stiffness. Blue dots in (**d**) and (**e**) are the configurations of the deformed lattice in (**a**) and (**b**). (**f**) Edge stiffness of hard and soft edges versus the stiffness, $k_2$, for multiple rest lengths of springs. As $k_2$ increases, the stiffnesses of the top and bottom edges tend to converge. (**g**) Elastic (solid) and shear (dashed) stiffness versus the twisting angle $\alpha$.
1212

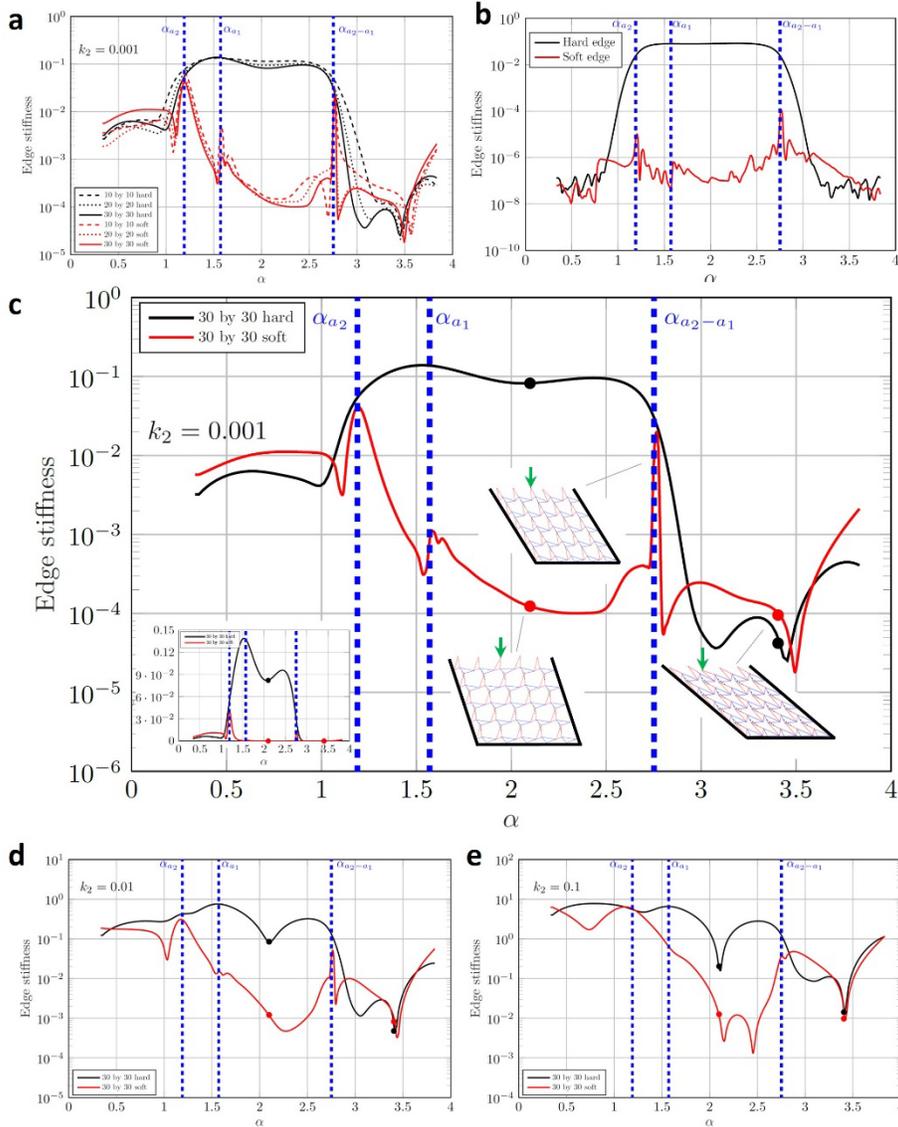

**Figure 3.** Linearized edge stiffness of kagome lattices transitioning between polarized and nonpolarized phases. (**a**) Edge stiffness of top and bottom edges from 10 × 10, 20 × 20 and 30 × 30 kagome lattices with structures and dimensions shown in Fig. 1. The rest length of the spring is set as 95% of $(a_r + b_b)/2$, and the lattice is held at different configurations (labelled by angle $\alpha$) by three boundaries with the edge stiffness measured at the fourth boundary. The bond stiffness $k_1$ and the spring constant $k_2$ are 1 and 0.001, respectively. (**b**) Edge stiffness of a 30 × 30 kagome lattice without bistable units ($k_2 = 0$). (**c-e**) Edge stiffness of a 30 × 30 kagome lattice with the same $k_1$ and different $k_2$. (**c**) $k_2 = 0.001$, (**d**) $k_2 = 0.01$, and (**e**) $k_2 = 0.1$. Black and red dots indicate the specific surface stiffness at the equilibrium twisting angles, $\alpha_0$, where the spring is relaxed, one located in the topologically polarized phase and the other in the non-polarized phase. An inset in (**c**) is a linear scale plot of edge stiffness for $k_2 = 0.001$.

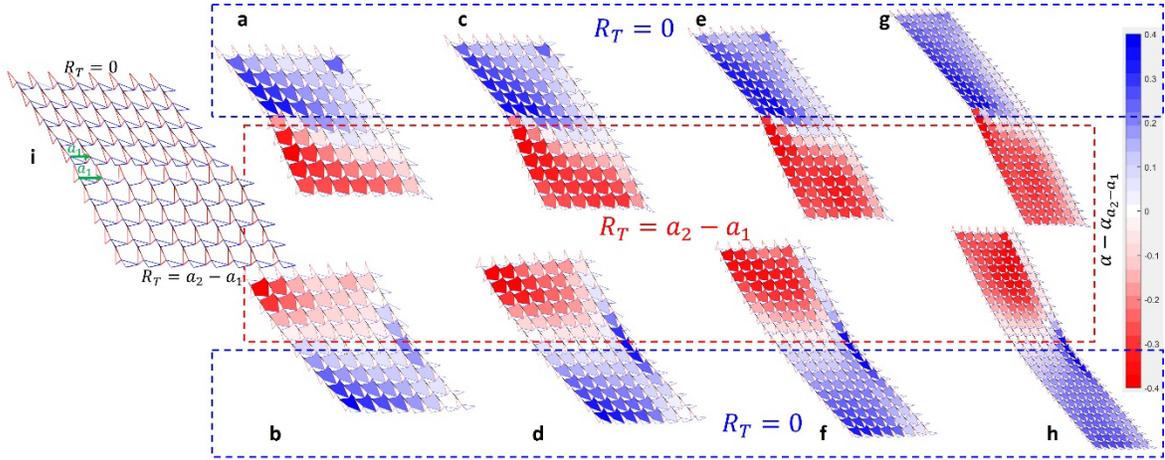

**Figure 4.** Reconfiguration interfaces of multistable topological kagome lattices. Reconfiguration interface of (**a-b**) 10 × 8, (**c-d**) 14 × 8, (**e-f**) 20 × 8, and (**g-h**) 30 × 8 kagome lattices based on minimization of nonlinear elastic energy of the lattice. (**a, c, e, g**) Initial configurations of the lattices have non-polarized domains at the top area of the lattice and polarized domains at the bottom area, respectively (forming a zero-mode interface). (**b, d, f, h**) Initial configuration presents that the non-polarized area is at the bottom while the polarized area is on the top part of the lattice, creating a self-stress-state interface. (**i**) Two uniform lattices with different primitive vectors are shown to illustrate fitting them together while with different widths. Examples of initial configurations are shown in Supplementary Materials, Figure S2. The colormap is used for the differences of the twisting angles α between unit cells, and $\alpha_{a_2-a_1}$—positive values (in blue) indicate non-polarized unit cells, and negative (in red) represent polarized unit cells. $\alpha - \alpha_{a_2-a_1} = \pm 0.2$ imply the unit cell is at its bistable equilibrium in which the spring is relaxed ($\alpha_0 = 2.95$ and $\alpha_0 = 2.55$).

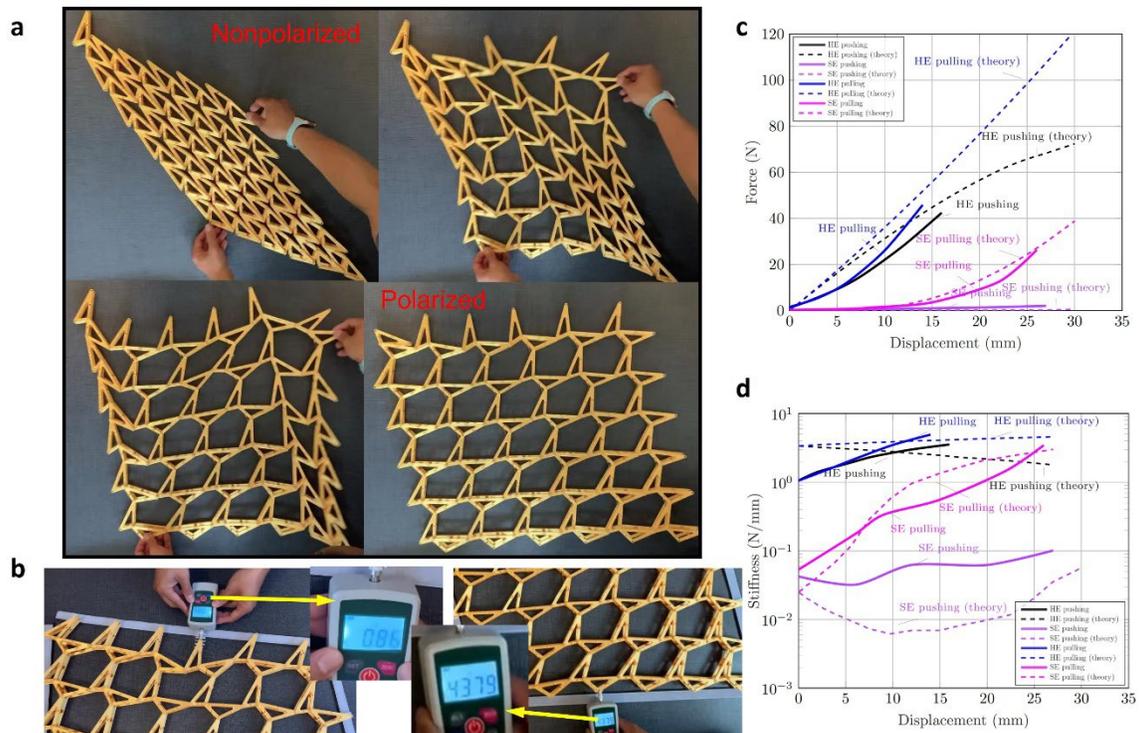

**Figure 5.** Experimental validation on topological transition and surface stiffness. (**a**) Topological transition (swift twist of angles) from non-polarized phase to polarized phase for a 6 × 6-unit cell lattice. (**b**) The comparison of forces needed to push the tips of two opposite edges at given displacements. The hard edge is pushed to 20 mm which needs over 43 N, while 19 mm is given for the soft edge needing only 0.86 N. (**c**) Force and (**d**) corresponding stiffness against displacement for two edges. The black (blue) line indicates the hard edge undergoes a pushing (pulling) load, while the red (magenta) represents soft edges applied by a pushing (pulling) force. Solid and dashed lines are used to distinguish theoretical results and experimental data.

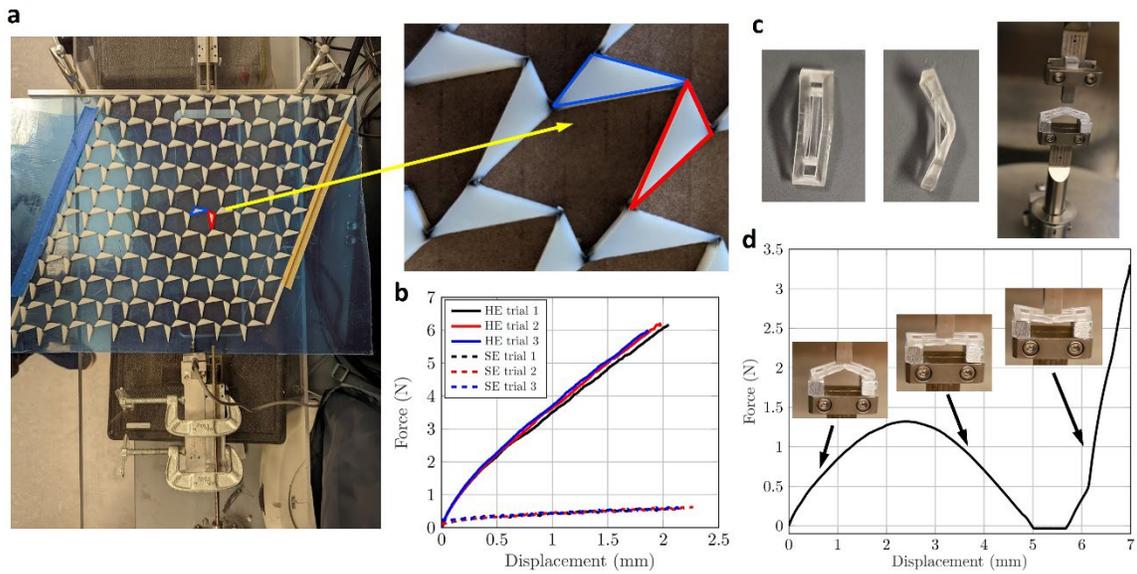

**Figure 6.** Surface stiffness of bistable lattice made by multi-materials and bistable unit design and test. (**a**) Experimental setup for surface stiffness testing of multi-material Maxwell lattice. (**b**) Force versus displacement for the hard and floppy edges. The solid lines indicate three trials for the hard edge while the dashed lines represent trials for the soft edge. The stiffness ratio between the two edges is averaged by the fitting with 3 trials and is around 16. (**c**) A bistable hinge, composed of two approximately triangular shaped prisms attached by a flexural hinge, and connected by a thin spring element. Experimental setup for force-displacement testing of the bistable hinge. (**d**) Force versus displacement for the bistable hinge. The force probe starts from one stable state of the bistable hinge and pushes the structure snapping to the other stable state with a displacement rate of 0.2 mm/s.

# Supplementary Text

## Note 1: Elastic energy of a unit cell in the lattice

A unit cell chosen to calculate the elastic energy is illustrated on Fig. 1a, where the displacement in $x$ and $y$ directions at each node has added to the unit cell, shown in Fig. S1. The bonds of the triangles can be defined as deformable rods with an effective stiffness, $k_1$, and the stiffness of the spring linking the centers of two bonds is $k_2$. When the unit cell is deformed associated with the lattice, displacements $\boldsymbol{u}$ of every nodes are generated, resulting in creating elastic energy of the unit cell.

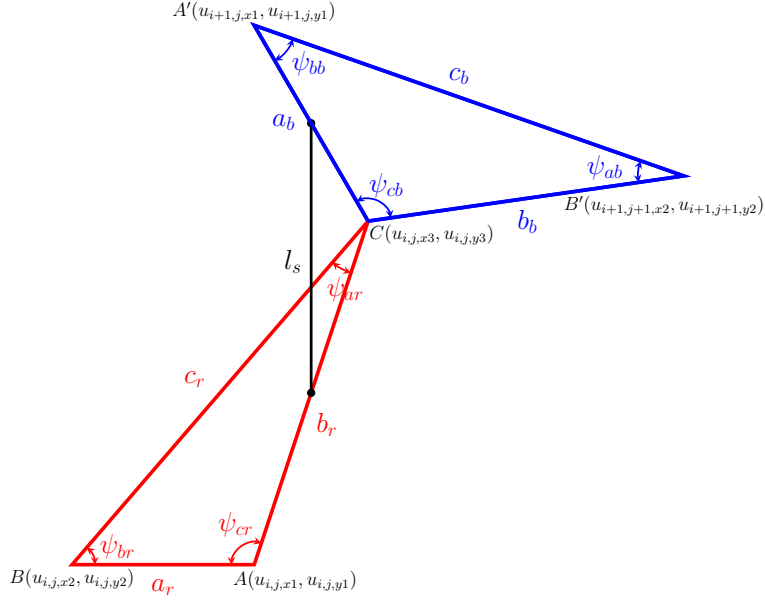

Figure S1: A unit cell of a homogeneous lattice with a bistable unit where the internal angle between two triangles is $\alpha$. Node $A$, $B$, $A'$, and $B'$ are shared with the adjacent unit cells, while Node $C$ belongs to the current unit cell corresponding to the $\alpha$ angle.

Nodes defined as $A$, $B$, $C$, $A'$, and $B'$ in the unit cell are shared by two triangles either from the same unit cell or from different units. To avoid double counting of these nodes, each unit cell only contains three nodes, $A$, $B$, and $C$, which have notations of $_{i,j}$, representing the unit cell from $i$th row and $j$th column of the lattice. Adding nodes from the boundaries, the total number of nodes for a lattice with $n \times m$ unit cells is $6nm + 2n + 2m$. The displacement vector of the unit cell $(i,j)$ shown in Fig. S1 is defined as

$$\boldsymbol{u} = \Big(u_{i,j,x1}, u_{i,j,x2}, u_{i,j,x3}, u_{i+1,j,x1}, u_{i+1,j+1,x2}, u_{i,j,y1}, u_{i,j,y2}, u_{i,j,y3}, u_{i+1,j,y1}, u_{i+1,j+1,y2}\Big)^T. \tag{S.1}$$

The total elastic energy in the unit cell consisting of the energy from six sides of triangles and the last one from the added

spring is given by

$$
\begin{aligned}
En = \frac{k_2}{2} & \left( \frac{1}{2}\sqrt{\begin{aligned}&[-u_{i,j,x1}+u_{i+1,j,x1}+b_r\cos(\theta+\psi_{bb}+\psi_{cr})-a_b\cos(\alpha+\theta+\psi_{ar}+\psi_{bb}+\psi_{cr})]^2\\&+[u_{i,j,y1}-u_{i+1,j,y1}+b_r\sin(\theta+\psi_{bb}+\psi_{cr})-a_b\sin(\alpha+\theta+\psi_{ar}+\psi_{bb}+\psi_{cr})]^2\end{aligned}}-l_s\right)^2\\
+\frac{k_1}{2}&\left[\left(\sqrt{\begin{aligned}&[-u_{i,j,x1}+u_{i,j,x2}+a_r\cos(\theta+\psi_{bb})]^2\\&+[u_{i,j,y1}-u_{i,j,y2}+a_r\sin(\theta+\psi_{bb})]^2\end{aligned}}-a_r\right)^2+\left(\sqrt{\begin{aligned}&[-u_{i,j,x1}+u_{i,j,x3}+b_r\cos(\theta+\psi_{bb}+\psi_{cr})]^2\\&+[u_{i,j,y1}-u_{i,j,y3}+b_r\sin(\theta+\psi_{bb}+\psi_{cr})]^2\end{aligned}}-b_r\right)^2\right.\\
+&\left(\sqrt{\begin{aligned}&[u_{i,j,x2}-u_{i,j,x3}+a_r\cos(\theta+\psi_{bb})-b_r\cos(\theta+\psi_{bb}+\psi_{cr})]^2\\&+[-u_{i,j,y2}+u_{i,j,y3}+a_r\sin(\theta+\psi_{bb})-b_r\sin(\theta+\psi_{bb}+\psi_{cr})]^2\end{aligned}}-c_r\right)^2\\
+&\left(\sqrt{\begin{aligned}&[-u_{i+1,j,x1}+u_{i,j,x3}+a_b\cos(\alpha+\theta+\psi_{ar}+\psi_{bb}+\psi_{cr})]^2\\&+[u_{i+1,j,y1}-u_{i,j,y3}+a_b\sin(\alpha+\theta+\psi_{ar}+\psi_{bb}+\psi_{cr})]^2\end{aligned}}-a_b\right)^2\\
+&\left(\sqrt{\begin{aligned}&[-u_{i+1,j+1,x2}+u_{i,j,x3}+b_b\cos(\alpha+\theta+\psi_{ar}+\psi_{bb}+\psi_{cr}+\psi_{cb})]^2\\&+[u_{i+1,j+1,y2}-u_{i,j,y3}+b_b\sin(\alpha+\theta+\psi_{ar}+\psi_{bb}+\psi_{cr}+\psi_{cb})]^2\end{aligned}}-b_b\right)^2\\
+&\left.\left(\sqrt{\begin{aligned}&[-u_{i+1,j,x1}+u_{i+1,j+1,x2}+a_b\cos(\alpha+\theta+\psi_{ar}+\psi_{bb}+\psi_{cr})-b_b\cos(\alpha+\theta+\psi_{ar}+\psi_{bb}+\psi_{cr}+\psi_{cb})]^2\\&+[u_{i+1,j,y1}-u_{i+1,j+1,y2}+a_b\sin(\alpha+\theta+\psi_{ar}+\psi_{bb}+\psi_{cr})-b_b\sin(\alpha+\theta+\psi_{ar}+\psi_{bb}+\psi_{cr}+\psi_{cb})]^2\end{aligned}}-c_b\right)^2\right].
\end{aligned}
\tag{S.2}
$$

Eq. S.2 gives the energy of an entire unit cell. However, the total energy of a lattice is the combination of these entire unit cells having added spring and additional unit cells which only include red triangles from the top edge or blue triangles from the bottom. These incomplete unit cells only have partial terms of energy in Eq. S.2 which are related to those triangles. Reconfiguration results in Fig. 4 are solved by nonlinear minimization of the total elastic energy of the lattice using Eq. S.2. First, two initial configurations of the lattice are given by two different sets of angles from homogeneous lattices, one located in the polarized domain, and the other in the non-polarized phase. Second, bonds at the interface of two groups of lattices are stretched and connected together to create an entire lattice consisting of these two domains (two examples in Fig. S2). Third, the combined lattice is relaxed and the nonlinear minimization (nonlinear conjugate gradient method) is applied to solve this unconstrained problem.

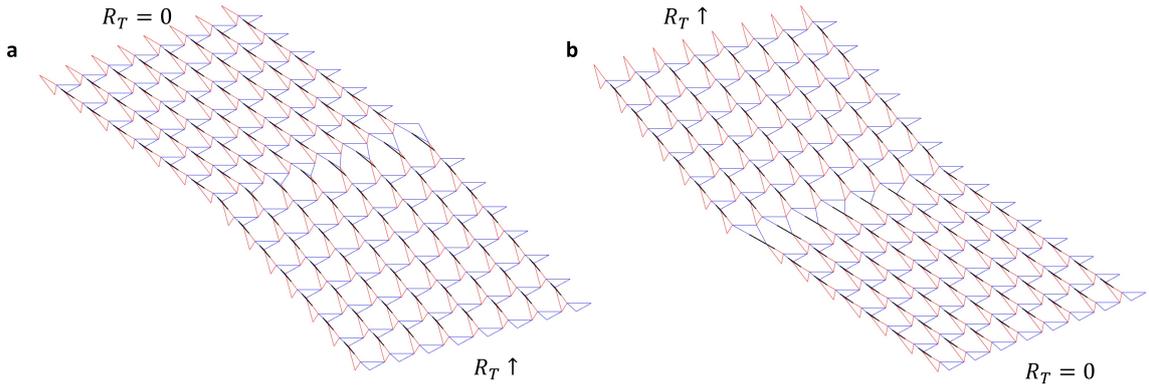

Figure S2: initial configurations of a $14 \times 8$ lattice with (a) $\boldsymbol{R}_T = \boldsymbol{a_2} - \boldsymbol{a_1}$ at the bottom and $\boldsymbol{R}_T = 0$ at the top, and (b) $\boldsymbol{R}_T = \boldsymbol{a_2} - \boldsymbol{a_1}$ at the top and $\boldsymbol{R}_T = 0$ at the bottom.

## Note 2: Linearized elastic energy of a unit cell

To quantify the surface stiffness of a Maxwell lattice during the polarization transition, it is assumed that the relative deformations of neighboring sites of the lattice is small enough (small $\Delta \boldsymbol{u}$ values) when compared with the geometry of the lattice, and thus the simplified linearized energy of a unit cell, $T_{en}$, can be used as an approximation instead of the nonlinear energy $En$. The linearized elastic energy of an entire unit cell represented in Fig. S1 is given by

$$T_{en} = \boldsymbol{u}^T \boldsymbol{K}_{en} \boldsymbol{u} + \boldsymbol{F}_{en}^T \boldsymbol{u} + T_{en0}, \tag{S.3}$$

where $\boldsymbol{K}_{en}$ and $\boldsymbol{F}_{en}$ are a 10 by 10 semi positive-definite matrix corresponding to the quadratic term and an 10 by 1 vector relate to the force on the bistable spring away from the rest length $l_s$, respectively, at a given $\alpha$, which have the forms

$$\boldsymbol{K}_{en} = \begin{bmatrix} K_{1,1} & K_{1,2} & K_{1,3} & K_{1,4} & 0 & K_{1,6} & K_{1,7} & K_{1,8} & K_{1,9} & 0 \\ K_{2,1} & K_{2,2} & K_{2,3} & 0 & 0 & K_{2,6} & K_{2,7} & K_{2,8} & 0 & 0 \\ K_{3,1} & K_{3,2} & K_{3,3} & K_{3,4} & K_{3,5} & K_{3,6} & K_{3,7} & K_{3,8} & K_{3,9} & K_{3,10} \\ K_{4,1} & 0 & K_{4,3} & K_{4,4} & K_{4,5} & K_{4,6} & 0 & K_{4,8} & K_{4,9} & K_{4,10} \\ 0 & 0 & K_{5,3} & K_{5,4} & K_{5,5} & 0 & 0 & K_{5,8} & K_{5,9} & K_{5,10} \\ K_{6,1} & K_{6,2} & K_{6,3} & K_{6,4} & 0 & K_{6,6} & K_{6,7} & K_{6,8} & K_{6,9} & 0 \\ K_{7,1} & K_{7,2} & K_{7,3} & 0 & 0 & K_{7,6} & K_{7,7} & K_{7,8} & 0 & 0 \\ K_{8,1} & K_{8,2} & K_{8,3} & K_{8,4} & K_{8,5} & K_{8,6} & K_{8,7} & K_{8,8} & K_{8,9} & K_{8,10} \\ K_{9,1} & 0 & K_{9,3} & K_{9,4} & K_{9,5} & K_{9,6} & 0 & K_{9,8} & K_{9,9} & K_{9,10} \\ 0 & 0 & K_{10,3} & K_{10,4} & K_{10,5} & 0 & 0 & K_{10,8} & K_{10,9} & K_{10,10} \end{bmatrix}_\alpha, \tag{S.4}$$

$$\boldsymbol{F}_{en}^T = \begin{bmatrix} F_1 & 0 & 0 & F_4 & 0 & F_6 & 0 & 0 & F_9 & 0 \end{bmatrix}_\alpha, \tag{S.5}$$

and $T_{en0} = \frac{k_2}{4}\left(a_b^2 + b_r^2 - 2a_b b_r \cos(\alpha + \psi_{ar}) + 4l_s(l_s - \sqrt{a_b^2 + b_r^2 - 2a_b b_r \cos(\alpha + \psi_{ar})})\right)$ is an energy constant resulting from the prestress in the spring, where $l_s$ is the rest length of the added spring.

Coefficients of diagonal terms in the dynamic matrix $\boldsymbol{K}_{en}$ are calculated:

$$K_{1,1} = k_1 \cos^2(\theta + \psi_{bb}) + k_1 \cos^2(\theta + \psi_{bb} + \psi_{cr}) + \frac{k_2(b_r \cos(\theta + \psi_{bb} + \psi_{cr}) - a_b \cos(\alpha + \theta + \psi_{ar} + \psi_{bb} + \psi_{cr}))^2}{4(a_b^2 + b_r^2 - 2a_b b_r \cos(\alpha + \psi_{ar}))} \tag{S.6}$$

$$K_{2,2} = k_1 \cos^2(\theta + \psi_{bb}) + \frac{k_1(a_r \cos(\theta + \psi_{bb}) - b_r \cos(\theta + \psi_{bb} + \psi_{cr}))^2}{a_r^2 + b_r^2 - 2a_r b_r \cos(\psi_{cr})} \tag{S.7}$$

$$\begin{aligned} K_{3,3} =& k_1 \cos^2(\theta + \psi_{bb} + \psi_{cr}) + k_1 \cos^2(\alpha + \theta + \psi_{ar} + \psi_{bb} + \psi_{cr}) + \\ & k_1 \cos^2(\alpha + \theta + \psi_{ar} + \psi_{bb} + \psi_{cr} + \psi_{cb}) + \frac{k_1(a_r \cos(\theta + \psi_{bb}) - b_r \cos(\theta + \psi_{bb} + \psi_{cr}))^2}{a_r^2 + b_r^2 - 2a_r b_r \cos(\psi_{cr})} \end{aligned} \tag{S.8}$$

$$\begin{aligned} K_{4,4} =& k_1 \cos^2(\alpha + \theta + \psi_{ar} + \psi_{bb} + \psi_{cr}) + \frac{k_2(b_r \cos(\theta + \psi_{bb} + \psi_{cr}) - a_b \cos(\alpha + \theta + \psi_{ar} + \psi_{bb} + \psi_{cr}))^2}{4(a_b^2 + b_r^2 - 2a_b b_r \cos(\alpha + \psi_{ar}))} + \\ & \frac{k_1(a_b \cos(\alpha + \theta + \psi_{ar} + \psi_{bb} + \psi_{cr}) - b_b \cos(\alpha + \theta + \psi_{ar} + \psi_{bb} + \psi_{cr} + \psi_{cb}))^2}{a_b^2 + b_b^2 - 2a_b b_b \cos(\psi_{cb})} \end{aligned} \tag{S.9}$$

$$\begin{aligned} K_{5,5} =& k_1 \cos^2(\alpha + \theta + \psi_{ar} + \psi_{bb} + \psi_{cr} + \psi_{cb}) + \\ & \frac{k_1(a_b \cos(\alpha + \theta + \psi_{ar} + \psi_{bb} + \psi_{cr}) - b_b \cos(\alpha + \theta + \psi_{ar} + \psi_{bb} + \psi_{cr} + \psi_{cb}))^2}{a_b^2 + b_b^2 - 2a_b b_b \cos(\psi_{cb})} \end{aligned} \tag{S.10}$$

$$K_{6,6} = k_1 \sin^2(\theta + \psi_{bb}) + k_1 \sin^2(\theta + \psi_{bb} + \psi_{cr}) + \frac{k_2(b_r \sin(\theta + \psi_{bb} + \psi_{cr}) - a_b \sin(\alpha + \theta + \psi_{ar} + \psi_{bb} + \psi_{cr}))^2}{4(a_b^2 + b_r^2 - 2a_b b_r \cos(\alpha + \psi_{ar}))} \tag{S.11}$$

$$\begin{aligned} K_{7,7} =& k_1 \sin^2(\theta + \psi_{bb}) + \\ & \frac{k_1(a_r \sin(\theta + \psi_{bb}) - b_r \sin(\theta + \psi_{bb} + \psi_{cr}))^2}{a_r^2 + b_r^2 - 2a_r b_r \cos(\psi_{cr})} \end{aligned} \tag{S.12}$$

$$\begin{aligned} K_{8,8} =& k_1 \sin^2(\theta + \psi_{bb} + \psi_{cr}) + k_1 \sin^2(\alpha + \theta + \psi_{ar} + \psi_{bb} + \psi_{cr}) + \\ & k_1 \sin^2(\alpha + \theta + \psi_{ar} + \psi_{bb} + \psi_{cr} + \psi_{cb}) + \frac{k_1(a_r \sin(\theta + \psi_{bb}) - b_r \sin(\theta + \psi_{bb} + \psi_{cr}))^2}{a_r^2 + b_r^2 - 2a_r b_r \cos(\psi_{cr})} \end{aligned} \tag{S.13}$$

$$\begin{aligned} K_{9,9} =& k_1 \sin^2(\alpha + \theta + \psi_{ar} + \psi_{bb} + \psi_{cr}) + \frac{k_2(b_r \sin(\theta + \psi_{bb} + \psi_{cr}) - a_b \sin(\alpha + \theta + \psi_{ar} + \psi_{bb} + \psi_{cr}))^2}{4(a_b^2 + b_r^2 - 2a_b b_r \cos(\alpha + \psi_{ar}))} + \\ & \frac{k_1(a_b \sin(\alpha + \theta + \psi_{ar} + \psi_{bb} + \psi_{cr}) - b_b \sin(\alpha + \theta + \psi_{ar} + \psi_{bb} + \psi_{cr} + \psi_{cb}))^2}{a_b^2 + b_b^2 - 2a_b b_b \cos(\psi_{cb})} \end{aligned} \tag{S.14}$$

$$K_{10,10} = k_1 \sin^2(\alpha + \theta + \psi_{ar} + \psi_{bb} + \psi_{cr} + \psi_{cb}) + \frac{k_1(a_b \sin(\alpha + \theta + \psi_{ar} + \psi_{bb} + \psi_{cr}) - b_b \sin(\alpha + \theta + \psi_{ar} + \psi_{bb} + \psi_{cr} + \psi_{cb}))^2}{a_b^2 + b_b^2 - 2a_b b_b \cos(\psi_{cb})} \cdot \tag{S.15}$$

Coefficients of off-diagonal terms in Eq. (S.4) are expressed below:

$$K_{1,2} = K_{2,1} = -k_1 \cos^2(\theta + \psi_{bb}) \tag{S.16}$$

$$K_{1,3} = K_{3,1} = -k_1 \cos^2(\theta + \psi_{bb} + \psi_{cr}) \tag{S.17}$$

$$K_{1,4} = K_{4,1} = \frac{-k_2(b_r \cos(\theta + \psi_{bb} + \psi_{cr}) - a_b \cos(\alpha + \theta + \psi_{ar} + \psi_{bb} + \psi_{cr}))^2}{4(a_b^2 + b_r^2 - 2a_b b_r \cos(\alpha + \psi_{ar}))} \tag{S.18}$$

$$K_{1,6} = K_{6,1} = -k_1 \cos(\theta + \psi_{bb}) \sin(\theta + \psi_{bb}) - k_1 \cos(\theta + \psi_{bb} + \psi_{cr}) \sin(\theta + \psi_{bb} + \psi_{cr}) - \\ k_2(b_r \cos(\theta + \psi_{bb} + \psi_{cr}) - a_b \cos(\alpha + \theta + \psi_{ar} + \psi_{bb} + \psi_{cr})) \cdot \\ \frac{(b_r \sin(\theta + \psi_{bb} + \psi_{cr}) - a_b \sin(\alpha + \theta + \psi_{ar} + \psi_{bb} + \psi_{cr}))}{4(a_b^2 + b_r^2 - 2a_b b_r \cos(\alpha + \psi_{ar}))} \tag{S.19}$$

$$K_{1,7} = K_{7,1} = k_1 \cos(\theta + \psi_{bb}) \sin(\theta + \psi_{bb}) \tag{S.20}$$

$$K_{1,8} = K_{8,1} = k_1 \cos(\theta + \psi_{bb} + \psi_{cr}) \sin(\theta + \psi_{bb} + \psi_{cr}) \tag{S.21}$$

$$K_{1,9} = K_{9,1} = k_2(b_r \cos(\theta + \psi_{bb} + \psi_{cr}) - a_b \cos(\alpha + \theta + \psi_{ar} + \psi_{bb} + \psi_{cr})) \cdot \\ \frac{(b_r \sin(\theta + \psi_{bb} + \psi_{cr}) - a_b \sin(\alpha + \theta + \psi_{ar} + \psi_{bb} + \psi_{cr}))}{4(a_b^2 + b_r^2 - 2a_b b_r \cos(\alpha + \psi_{ar}))} \tag{S.22}$$

$$K_{2,3} = K_{3,2} = \frac{-k_1(a_r \cos(\theta + \psi_{bb}) - b_r \cos(\theta + \psi_{bb} + \psi_{cr}))^2}{a_r^2 + b_r^2 - 2a_r b_r \cos(\psi_{cr})} \tag{S.23}$$

$$K_{2,6} = K_{6,2} = k_1 \cos(\theta + \psi_{bb}) \sin(\theta + \psi_{bb}) \tag{S.24}$$

$$K_{2,7} = K_{7,2} = k_1 \cos(\theta + \psi_{bb}) \sin(\theta + \psi_{bb}) + \\ \frac{k_1(a_r \cos(\theta + \psi_{bb}) - b_r \cos(\theta + \psi_{bb} + \psi_{cr}))(-a_r \sin(\theta + \psi_{bb}) + b_r \sin(\theta + \psi_{bb} + \psi_{cr}))}{a_r^2 + b_r^2 - 2a_r b_r \cos(\psi_{cr})} \tag{S.25}$$

$$K_{2,8} = K_{8,2} = \frac{k_1(a_r \cos(\theta + \psi_{bb}) - b_r \cos(\theta + \psi_{bb} + \psi_{cr}))(a_r \sin(\theta + \psi_{bb}) - b_r \sin(\theta + \psi_{bb} + \psi_{cr}))}{a_r^2 + b_r^2 - 2a_r b_r \cos(\psi_{cr})} \tag{S.26}$$

$$K_{3,4} = K_{4,3} = -k_1 \cos^2(\alpha + \theta + \psi_{ar} + \psi_{bb} + \psi_{cr}) \tag{S.27}$$

$$K_{3,5} = K_{5,3} = -k_1 \cos^2(\alpha + \theta + \psi_{ar} + \psi_{bb} + \psi_{cr} + \psi_{cb}) \tag{S.28}$$

$$K_{3,6} = K_{6,3} = k_1 \cos(\theta + \psi_{bb} + \psi_{cr}) \sin(\theta + \psi_{bb} + \psi_{cr}) \tag{S.29}$$

$$K_{3,7} = K_{7,3} = \frac{k_1(a_r \cos(\theta + \psi_{bb}) - b_r \cos(\theta + \psi_{bb} + \psi_{cr}))(a_r \sin(\theta + \psi_{bb}) - b_r \sin(\theta + \psi_{bb} + \psi_{cr}))}{a_r^2 + b_r^2 - 2a_r b_r \cos(\psi_{cr})} \tag{S.30}$$

$$K_{3,8} = K_{8,3} = -k_1 \cos(\theta + \psi_{bb} + \psi_{cr}) \sin(\theta + \psi_{bb} + \psi_{cr}) + \\ - k_1 \cos(\alpha + \theta + \psi_{ar} + \psi_{bb} + \psi_{cr}) \sin(\alpha + \theta + \psi_{ar} + \psi_{bb} + \psi_{cr}) + \\ - k_1 \cos(\alpha + \theta + \psi_{ar} + \psi_{bb} + \psi_{cr} + \psi_{cb}) \sin(\alpha + \theta + \psi_{ar} + \psi_{bb} + \psi_{cr} + \psi_{cb}) + \\ \frac{k_1(a_r \cos(\theta + \psi_{bb}) - b_r \cos(\theta + \psi_{bb} + \psi_{cr}))(-a_r \sin(\theta + \psi_{bb}) + b_r \sin(\theta + \psi_{bb} + \psi_{cr}))}{a_r^2 + b_r^2 - 2a_r b_r \cos(\psi_{cr})} \tag{S.31}$$

$$K_{3,9} = K_{9,3} = k_1 \cos(\alpha + \theta + \psi_{ar} + \psi_{bb} + \psi_{cr}) \sin(\alpha + \theta + \psi_{ar} + \psi_{bb} + \psi_{cr}) \tag{S.32}$$

$$K_{3,10} = K_{10,3} = k_1 \cos(\alpha + \theta + \psi_{ar} + \psi_{bb} + \psi_{cr} + \psi_{cb}) \sin(\alpha + \theta + \psi_{ar} + \psi_{bb} + \psi_{cr} + \psi_{cb}) \tag{S.33}$$

$$K_{4,5} = K_{5,4} = \frac{-k_1(a_b \cos(\alpha + \theta + \psi_{ar} + \psi_{bb} + \psi_{cr}) - b_b \cos(\alpha + \theta + \psi_{ar} + \psi_{bb} + \psi_{cr} + \psi_{cb}))^2}{a_b^2 + b_b^2 - 2a_b b_b \cos(\psi_{cb})} \tag{S.34}$$

$$K_{4,6} = K_{6,4} = k_2(b_r \cos(\theta + \psi_{bb} + \psi_{cr}) - a_b \cos(\alpha + \theta + \psi_{ar} + \psi_{bb} + \psi_{cr})) \cdot \\ \frac{(b_r \sin(\theta + \psi_{bb} + \psi_{cr}) - a_b \sin(\alpha + \theta + \psi_{ar} + \psi_{bb} + \psi_{cr}))}{4(a_b^2 + b_r^2 - 2a_b b_r \cos(\alpha + \psi_{ar}))} \tag{S.35}$$

$$K_{4,8} = K_{8,4} = k_1 \cos(\alpha + \theta + \psi_{ar} + \psi_{bb} + \psi_{cr}) \sin(\alpha + \theta + \psi_{ar} + \psi_{bb} + \psi_{cr}) \tag{S.36}$$

$$\begin{aligned} K_{4,9} = K_{9,4} = &-k_1 \cos(\alpha + \theta + \psi_{ar} + \psi_{bb} + \psi_{cr}) \sin(\alpha + \theta + \psi_{ar} + \psi_{bb} + \psi_{cr}) + \\ &- k_2(b_r \cos(\theta + \psi_{bb} + \psi_{cr}) - a_b \cos(\alpha + \theta + \psi_{ar} + \psi_{bb} + \psi_{cr})) \cdot \\ &\frac{(b_r \sin(\theta + \psi_{bb} + \psi_{cr}) - a_b \sin(\alpha + \theta + \psi_{ar} + \psi_{bb} + \psi_{cr}))}{4(a_b^2 + b_r^2 - 2a_b b_r \cos(\alpha + \psi_{ar}))} + \\ &- k_1(a_b \cos(\alpha + \theta + \psi_{ar} + \psi_{bb} + \psi_{cr}) - b_b \cos(\alpha + \theta + \psi_{ar} + \psi_{bb} + \psi_{cr} + \psi_{cb})) \cdot \\ &\frac{(a_b \sin(\alpha + \theta + \psi_{ar} + \psi_{bb} + \psi_{cr}) - b_b \sin(\alpha + \theta + \psi_{ar} + \psi_{bb} + \psi_{cr} + \psi_{cb}))}{a_b^2 + b_b^2 - 2a_b b_b \cos(\psi_{cb})} \end{aligned} \tag{S.37}$$

$$\begin{aligned} K_{4,10} = K_{10,4} = &k_1(a_b \cos(\alpha + \theta + \psi_{ar} + \psi_{bb} + \psi_{cr}) - b_b \cos(\alpha + \theta + \psi_{ar} + \psi_{bb} + \psi_{cr} + \psi_{cb})) \cdot \\ &\frac{(a_b \sin(\alpha + \theta + \psi_{ar} + \psi_{bb} + \psi_{cr}) - b_b \sin(\alpha + \theta + \psi_{ar} + \psi_{bb} + \psi_{cr} + \psi_{cb}))}{a_b^2 + b_b^2 - 2a_b b_b \cos(\psi_{cb})} \end{aligned} \tag{S.38}$$

$$K_{5,8} = K_{8,5} = k_1 \cos(\alpha + \theta + \psi_{ar} + \psi_{bb} + \psi_{cr} + \psi_{cb}) \sin(\alpha + \theta + \psi_{ar} + \psi_{bb} + \psi_{cr} + \psi_{cb}) \tag{S.39}$$

$$\begin{aligned} K_{5,9} = K_{9,5} = &k_1(a_b \cos(\alpha + \theta + \psi_{ar} + \psi_{bb} + \psi_{cr}) - b_b \cos(\alpha + \theta + \psi_{ar} + \psi_{bb} + \psi_{cr} + \psi_{cb})) \cdot \\ &\frac{(a_b \sin(\alpha + \theta + \psi_{ar} + \psi_{bb} + \psi_{cr}) - b_b \sin(\alpha + \theta + \psi_{ar} + \psi_{bb} + \psi_{cr} + \psi_{cb}))}{a_b^2 + b_b^2 - 2a_b b_b \cos(\psi_{cb})} \end{aligned} \tag{S.40}$$

$$\begin{aligned} K_{5,10} = K_{10,5} = &-k_1 \cos(\alpha + \theta + \psi_{ar} + \psi_{bb} + \psi_{cr} + \psi_{cb}) \sin(\alpha + \theta + \psi_{ar} + \psi_{bb} + \psi_{cr} + \psi_{cb}) - \\ &k_1(a_b \cos(\alpha + \theta + \psi_{ar} + \psi_{bb} + \psi_{cr}) - b_b \cos(\alpha + \theta + \psi_{ar} + \psi_{bb} + \psi_{cr} + \psi_{cb})) \cdot \\ &\frac{(a_b \sin(\alpha + \theta + \psi_{ar} + \psi_{bb} + \psi_{cr}) - b_b \sin(\alpha + \theta + \psi_{ar} + \psi_{bb} + \psi_{cr} + \psi_{cb}))}{a_b^2 + b_b^2 - 2a_b b_b \cos(\psi_{cb})} \end{aligned} \tag{S.41}$$

$$K_{6,7} = K_{7,6} = -k_1 \sin^2(\theta + \psi_{bb}) \tag{S.42}$$

$$K_{6,8} = K_{8,6} = -k_1 \sin^2(\theta + \psi_{bb} + \psi_{cr}) \tag{S.43}$$

$$K_{6,9} = K_{9,6} = \frac{-k_2(b_r \sin(\theta + \psi_{bb} + \psi_{cr}) - a_b \sin(\alpha + \theta + \psi_{ar} + \psi_{bb} + \psi_{cr}))^2}{4(a_b^2 + b_r^2 - 2a_b b_r \cos(\alpha + \psi_{ar}))} \tag{S.44}$$

$$K_{7,8} = K_{8,7} = \frac{-k_1(a_r \sin(\theta + \psi_{bb}) - b_r \sin(\theta + \psi_{bb} + \psi_{cr}))^2}{a_r^2 + b_r^2 - 2a_r b_r \cos(\psi_{cr})} \tag{S.45}$$

$$K_{8,9} = K_{9,8} = -k_1 \sin^2(\alpha + \theta + \psi_{ar} + \psi_{bb} + \psi_{cr}) \tag{S.46}$$

$$K_{8,10} = K_{10,8} = -k_1 \sin^2(\alpha + \theta + \psi_{ar} + \psi_{bb} + \psi_{cr} + \psi_{cb}) \tag{S.47}$$

$$K_{9,10} = K_{10,9} = \frac{-k_1(a_b \sin(\alpha + \theta + \psi_{ar} + \psi_{bb} + \psi_{cr}) - b_b \sin(\alpha + \theta + \psi_{ar} + \psi_{bb} + \psi_{cr} + \psi_{cb}))^2}{a_b^2 + b_b^2 - 2a_b b_b \cos(\psi_{cb})}. \tag{S.48}$$

Additional coefficients in the $\boldsymbol{F}_{en}$ vector are

$$\begin{aligned} F_1 = &k_2(-2l_s + \sqrt{a_b^2 + b_r^2 - 2a_b b_r \cos(\alpha + \psi_{ar})}) \cdot \\ &\frac{-b_r \cos(\alpha + \psi_{ar}) + a_b \cos(\alpha + \theta + \psi_{ar} + \psi_{bb} + \psi_{cr})}{2\sqrt{a_b^2 + b_r^2 - 2a_b b_r \cos(\alpha + \psi_{ar})}} \end{aligned} \tag{S.49}$$

$$\begin{aligned} F_4 = &k_2(-2l_s + \sqrt{a_b^2 + b_r^2 - 2a_b b_r \cos(\alpha + \psi_{ar})}) \cdot \\ &\frac{b_r \cos(\alpha + \psi_{ar}) - a_b \cos(\alpha + \theta + \psi_{ar} + \psi_{bb} + \psi_{cr})}{2\sqrt{a_b^2 + b_r^2 - 2a_b b_r \cos(\alpha + \psi_{ar})}} \end{aligned} \tag{S.50}$$

$$\begin{aligned} F_6 = &k_2(-2l_s + \sqrt{a_b^2 + b_r^2 - 2a_b b_r \cos(\alpha + \psi_{ar})}) \cdot \\ &\frac{b_r \sin(\alpha + \psi_{ar}) - a_b \sin(\alpha + \theta + \psi_{ar} + \psi_{bb} + \psi_{cr})}{2\sqrt{a_b^2 + b_r^2 - 2a_b b_r \cos(\alpha + \psi_{ar})}} \end{aligned} \tag{S.51}$$

$$F_9 = k_2(-2l_s + \sqrt{a_b^2 + b_r^2 - 2a_b b_r \cos(\alpha + \psi_{ar})}) \cdot$$
$$\frac{-b_r \sin(\alpha + \psi_{ar}) + a_b \sin(\alpha + \theta + \psi_{ar} + \psi_{bb} + \psi_{cr})}{2\sqrt{a_b^2 + b_r^2 - 2a_b b_r \cos(\alpha + \psi_{ar})}} \cdot \quad \text{(S.52)}$$

The lattice also consists of unit cells with only red or blue triangles at the top or bottom edges without bistable units. For a unit cell, $(1, j)$, which only includes a blue triangle at the bottom of the lattice, the displacement vector $\boldsymbol{u}$ only consists of Node $A'$, $B'$, and $C$ in Fig. S1, and is represented as

$$\boldsymbol{u} = \left( u_{i+1,j,x1}, u_{i+1,j+1,x2}, u_{i,j,x3}, u_{i+1,j,y1}, u_{i+1,j+1,y2}, u_{i,j,y3} \right)^T. \quad \text{(S.53)}$$

The corresponding linearized energy can be written by

$$T_b = \boldsymbol{u}^T \boldsymbol{K}_b \boldsymbol{u}, \quad \text{(S.54)}$$

where $\boldsymbol{K}_b$ is a 6 by 6 semi positive-definite matrix and is expressed

$$\boldsymbol{K}_b = \begin{bmatrix} Kb_{1,1} & Kb_{1,2} & Kb_{1,3} & Kb_{1,4} & Kb_{1,5} & Kb_{1,6} \\ Kb_{2,1} & Kb_{2,2} & Kb_{2,3} & Kb_{2,4} & Kb_{2,5} & Kb_{2,6} \\ Kb_{3,1} & Kb_{3,2} & Kb_{3,3} & Kb_{3,4} & Kb_{3,5} & Kb_{3,6} \\ Kb_{4,1} & Kb_{4,2} & Kb_{4,3} & Kb_{4,4} & Kb_{4,5} & Kb_{4,6} \\ Kb_{5,1} & Kb_{5,2} & Kb_{5,3} & Kb_{5,4} & Kb_{5,5} & Kb_{5,6} \\ Kb_{6,1} & Kb_{6,2} & Kb_{6,3} & Kb_{6,4} & Kb_{6,5} & Kb_{6,6} \end{bmatrix}. \quad \text{(S.55)}$$

Each coefficient in the $\boldsymbol{K}_b$ is given as follow

$$Kb_{1,1} = k_1 + k_1 \cos^2(\psi_{bb}) \quad Kb_{2,2} = k_1 + \frac{k_1(c_b - a_b \cos(\psi_{bb}))^2}{a_b^2 + c_b^2 - 2a_b c_b \cos(\psi_{bb})} \quad Kb_{3,3} = k_1 \cos^2(\psi_{bb}) + \frac{k_1(c_b - a_b \cos(\psi_{bb}))^2}{a_b^2 + c_b^2 - 2a_b c_b \cos(\psi_{bb})} \quad \text{(S.56)}$$

$$Kb_{4,4} = k_1 \sin^2(\psi_{bb}) \quad Kb_{5,5} = \frac{k_1(a_b \sin(\psi_{bb}))^2}{a_b^2 + c_b^2 - 2a_b c_b \cos(\psi_{bb})} \quad Kb_{6,6} = k_1 \sin^2(\psi_{bb}) + \frac{k_1(a_b \sin(\psi_{bb}))^2}{a_b^2 + c_b^2 - 2a_b c_b \cos(\psi_{bb})}, \quad \text{(S.57)}$$

$$Kb_{1,2} = Kb_{2,1} = -k_1 \qquad Kb_{1,3} = Kb_{3,1} = -k_1 \cos^2(\psi_{bb}) \quad \text{(S.58)}$$

$$Kb_{1,4} = Kb_{4,1} = -k_1 \cos(\psi_{bb}) \sin(\psi_{bb}) \qquad Kb_{1,6} = Kb_{6,1} = k_1 \cos(\psi_{bb}) \sin(\psi_{bb}) \quad \text{(S.59)}$$

$$Kb_{2,3} = Kb_{3,2} = -\frac{k_1(c_b - a_b \cos(\psi_{bb}))^2}{a_b^2 + c_b^2 - 2a_b c_b \cos(\psi_{bb})} \qquad Kb_{2,5} = Kb_{5,2} = \frac{k_1(c_b - a_b \cos(\psi_{bb}))(a_b \sin(\psi_{bb}))}{a_b^2 + c_b^2 - 2a_b c_b \cos(\psi_{bb})} \quad \text{(S.60)}$$

$$Kb_{2,6} = Kb_{6,2} = \frac{-k_1(c_b - a_b \cos(\psi_{bb}))(a_b \sin(\psi_{bb}))}{a_b^2 + c_b^2 - 2a_b c_b \cos(\psi_{bb})} \qquad Kb_{3,4} = Kb_{4,3} = k_1 \cos(\psi_{bb}) \sin(\psi_{bb}) \quad \text{(S.61)}$$

$$Kb_{3,5} = Kb_{5,3} = \frac{-k_1(c_b - a_b \cos(\psi_{bb}))(a_b \sin(\psi_{bb}))}{a_b^2 + c_b^2 - 2a_b c_b \cos(\psi_{bb})} \qquad \begin{aligned} Kb_{3,6} = Kb_{6,3} = &-k_1 \cos(\psi_{bb}) \sin(\psi_{bb}) + \\ & \frac{k_1(c_b - a_b \cos(\psi_{bb}))(a_b \sin(\psi_{bb}))}{a_b^2 + c_b^2 - 2a_b c_b \cos(\psi_{bb})} \end{aligned} \quad \text{(S.62)}$$

$$Kb_{4,6} = Kb_{6,4} = -k_1 \sin^2(\psi_{bb}) \qquad Kb_{5,6} = Kb_{6,5} = \frac{-k_1(a_b \sin(\psi_{bb}))^2}{a_b^2 + c_b^2 - 2a_b c_b \cos(\psi_{bb})}. \quad \text{(S.63)}$$

For a unit cell, $(n, j)$, that only has a red triangle at the top edge, the displacement vector $\boldsymbol{u}$ now has Nodes $A$, $B$, and $C$ accordingly. The corresponding semi positive-definite matrix $\boldsymbol{K}_r$ of this unit cell associated with these displacements is shown below.

$$\boldsymbol{u} = \left( u_{i,j,x1}, u_{i,j,x2}, u_{i,j,x3}, u_{i,j,y1}, u_{i,j,y2}, u_{i,j,y3} \right)^T, \quad \text{(S.64)}$$

$$\boldsymbol{K}_r = \begin{bmatrix} K_{1,1r} & K_{1,2} & K_{1,3} & K_{1,6r} & K_{1,7} & K_{1,8} \\ K_{2,1} & K_{2,2} & K_{2,3} & K_{2,6} & K_{2,7} & K_{2,8} \\ K_{3,1} & K_{3,2} & K_{3,3r} & K_{3,6} & K_{3,7} & K_{3,8r} \\ K_{6,1r} & K_{6,2} & K_{6,3} & K_{6,6r} & K_{6,7} & K_{6,8} \\ K_{7,1} & K_{7,2} & K_{7,3} & K_{7,6} & K_{7,7} & K_{7,8} \\ K_{8,1} & K_{8,2} & K_{8,3r} & K_{8,6} & K_{8,7} & K_{8,8r} \end{bmatrix}_\alpha, \quad \text{(S.65)}$$

where $K_{i,j}$ is the coefficient from the dynamic matrix $\boldsymbol{K}_{en}$, and $K_{i,jr}$ are given by

$$K_{1,1r} = k_1 \cos^2(\theta + \psi_{bb}) + k_1 \cos^2(\theta + \psi_{bb} + \psi_{cr}) \tag{S.66}$$

$$K_{3,3r} = k_1 \cos^2(\theta + \psi_{bb} + \psi_{cr}) + \frac{k_1(a_r \cos(\theta + \psi_{bb}) - b_r \cos(\theta + \psi_{bb} + \psi_{cr}))^2}{a_r^2 + b_r^2 - 2a_r b_r \cos(\psi_{cr})} \tag{S.67}$$

$$K_{6,6r} = k_1 \sin^2(\theta + \psi_{bb}) + k_1 \sin^2(\theta + \psi_{bb} + \psi_{cr}) \tag{S.68}$$

$$K_{8,8r} = k_1 \sin^2(\theta + \psi_{bb} + \psi_{cr}) + \frac{k_1(-a_r \sin(\theta + \psi_{bb}) + b_r \sin(\theta + \psi_{bb} + \psi_{cr}))^2}{a_r^2 + b_r^2 - 2a_r b_r \cos(\psi_{cr})} \tag{S.69}$$

$$K_{1,6r} = K_{6,1r} = -k_1 \cos(\theta + \psi_{bb}) \sin(\theta + \psi_{bb}) - k_1 \cos(\theta + \psi_{bb} + \psi_{cr}) \sin(\theta + \psi_{bb} + \psi_{cr}) \tag{S.70}$$

$$\begin{aligned} K_{3,8r} = K_{8,3r} = &-k_1 \cos(\theta + \psi_{bb} + \psi_{cr}) \sin(\theta + \psi_{bb} + \psi_{cr}) - \\ & \frac{k_1(a_r \cos(\theta + \psi_{bb}) - b_r \cos(\theta + \psi_{bb} + \psi_{cr}))(a_r \sin(\theta + \psi_{bb}) - b_r \sin(\theta + \psi_{bb} + \psi_{cr}))}{a_r^2 + b_r^2 - 2a_r b_r \cos(\psi_{cr})}. \end{aligned} \tag{S.71}$$

Compared $K_{i,jr}$ with $K_{i,j}$ at the same position in $\boldsymbol{K}_{en}$, $K_{i,jr}$ does not include the terms related to the energy from the bistable unit.

The total linearized elastic energy, $T = \boldsymbol{u}^T \boldsymbol{K} \boldsymbol{u} + \boldsymbol{F}^T \boldsymbol{u} + T_0$, in Eq. (1) is the sum of the energy from all unit cells, including all those units with or without springs. The composite dynamic matrix $K$ is a linear combination of $\boldsymbol{K}_{en}$, $\boldsymbol{K}_b$ and $\boldsymbol{K}_r$ that forms a $N \times N$ matrix, where $N = 6nm + 2n + 2m$, and the composite vector $\boldsymbol{F}$ is a $N$ by 1 coefficient vector generated from $\boldsymbol{F}_{en}$ of each unit cell. The energy constant, $T_0 = (n-1)m T_{en0}$, is the sum of $T_{en0}$ from $(n-1)m$ unit cells which have bistable units.

## Note 3: Linearized mechanical stiffness

Without bistable units, a homogeneous Maxwell lattice can be transformed between different topological polarization domains via strain-free soft twisting. However, during the twisting process, the introduced bistability can generate new mechanical stiffness for the lattice, such as effective elastic and shear stiffness, when the spring is under compressed/elongated.

It is assumed that all bonds of the lattice have infinite stiffness and only added springs generate the elastic energy of the system. Starting from a homogeneous lattice with an initial twisting angle $\alpha$, the rest length of the added spring is

$$l_s = \sqrt{\frac{a_b^2}{4} + \frac{b_r^2}{4} - \frac{a_b b_r \cos(\alpha + \psi_{ar})}{2}}, \tag{S.72}$$

where the side length and inner angle of triangles are shown in Fig S1. Fixing the red triangle and given a small displacement, $u_x$, horizontal to the primitive vector, $\mathbf{a}_1$, (or $u_y$, perpendicular to $\mathbf{a}_1$) at Node $A'$, the resulting angle change at Node $C$ happens defined as $\delta\alpha$, and the corresponding length of the spring is

$$l = \sqrt{\frac{a_b^2}{4} + \frac{b_r^2}{4} - \frac{a_b b_r \cos(\alpha + \delta\alpha + \psi_{ar})}{2}}. \tag{S.73}$$

According to the geometric relation of two triangles in a unit cell, the small displacement $u_x$ and $u_y$ can be rewritten as a function of $\delta\alpha$:

$$\begin{aligned} u_x &= a_b \cos(\pi - \psi_{bb} - \delta\alpha) - a_b \cos(\pi - \psi_{bb}), \\ u_y &= a_b \sin(\pi - \psi_{bb} - \delta\alpha) - a_b \sin(\pi - \psi_{bb}). \end{aligned} \tag{S.74}$$

Since the displacement and angle change is assumed to be small enough, Eqs. (S.73-S.74) can be linearized and are given as:

$$l = l_s\left(1 + \frac{a_b b_r \sin(\alpha + \psi_{ar})}{2l_s^2}\delta\alpha\right), \tag{S.75}$$

$$\begin{aligned} u_x &= a_b \sin(\pi - \psi_{bb})\delta\alpha, \\ u_y &= -a_b \cos(\pi - \psi_{bb})\delta\alpha. \end{aligned} \tag{S.76}$$

The pushing force of $F_x$ ($F_y$) parallel (perpendicular) to the primitive vector of $\mathbf{a}_1$ that twists the lattice can be obtained by calculating the work that equals the total elastic energy of the added spring.

$$F_x = \frac{k_2}{2u_x}\delta l^2,$$
$$F_y = \frac{k_2}{2u_y}\delta l^2,$$
(S.77)

where $\delta l$ is the difference of the current length of spring and the rest length of the spring, which is

$$\delta l = l - l_s = \frac{a_b b_r \sin(\alpha + \psi_{ar})}{2l_s}\delta\alpha.$$
(S.78)

Therefore, the effective elastic stiffness and shear stiffness are

$$k_{el} = \frac{F_y}{u_y|\cos\phi|} = k_2 \frac{b_r^2 \sin^2(\alpha + \psi_{ar})}{8l_s^2 \cos^2(\pi - \psi_{bb})|\cos\phi|},$$
$$k_{sh} = \frac{F_x}{u_x|\sin\phi|} = k_2 \frac{b_r^2 \sin^2(\alpha + \psi_{ar})}{8l_s^2 \sin^2(\pi - \psi_{bb})|\sin\phi|},$$
(S.79)

where $\phi$ is the angle between the vector of $\mathbf{a}_1$ and the side of $c_b$, which is given by

$$\arctan\phi = \frac{a_r \sin(\alpha - \psi_{bb} - \psi_{br})}{c_b + a_r \cos(\alpha - \psi_{bb} - \psi_{br})}.$$
(S.80)

## Note 4: Information on the 3D printed Maxwell lattice

The prototype is designed using SOLIDWORKS and 3D printed using PLA materials via an Ender 3 Pro printer. The material infill density of 10% is used for the 3D printed parts. The spring is selected from commercial Uxewll 304 stainless steel, which has a spring constant of $k_2 = 150$ N/m. Fig. S3 shows a 3D-assembled designed unit cell, spring connectors and an example of an assembly of bistable lattice. Two rigid triangles (blue and red in Fig. S3a) are jointed at a vertex of each triangle to create a flexible hinge. Thus, the two triangles can be able to freely rotate to each other, so do the triangles in different unit cells. Moreover, the triangles are consisted of two layers, which gives the space for the additional spring embedded into the triangles. This added spring with much lower stiffness can provide an additional constraint of the unit and realize bistable equilibria (shown in Fig. 5a). The spring connectors (Fig. S3b) have raised parts which can hold the spring providing compressive or stretching force.

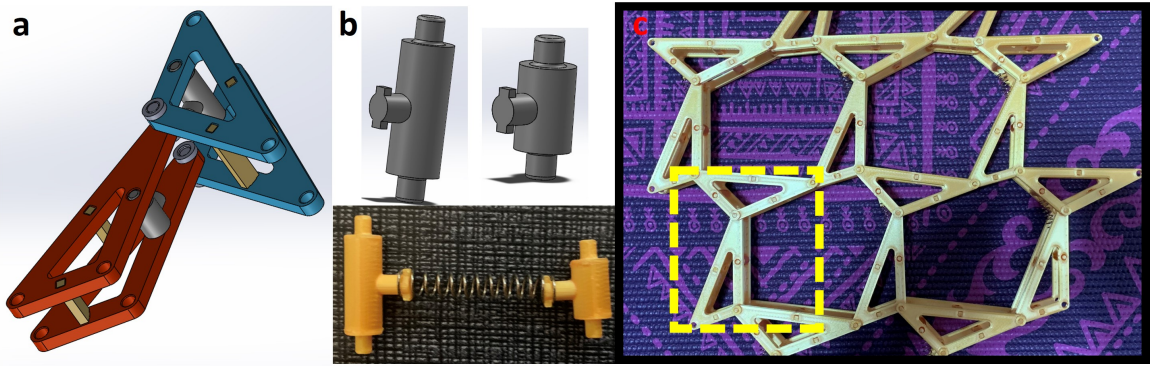

Figure S3: (a) 3-dimensional model of a unit cell, (b) 3D modeled and printed spring connectors, and (c) assembled unit cells connected with springs.

The effective stiffness $k_1$ for bonds of unit cells are determined by [1–3]

$$k_1 = \frac{f}{\Delta L} = 2E\frac{S}{L},$$
(S.81)

where $f$ is the stretching/compression force applied to the bond, $\Delta L$ is the stretched/compressed displacement, $E \approx 500$ MPa is the Young's modulus of the PLA material [4–6] (the infill density, printing speed and temperature, and holes in the part are expected to reduce the Young's modulus of PLA), and $S = 2.5 \times 10^{-6}$ m$^2$ and $L$ are the cross-sectional area and the length of the bond, respectively. The factor of 2 is due to the double-layered configuration of the printed unit cell. In the experiments, since the length of the each bond are different ($L = 0.09, 0.036, 0.072, 0.045,$ and $0.063$ m), $k_1$ is used to calculate the theoretical results with different values, which are shown as $k_1 = 2.778 \times 10^4, 6.944 \times 10^4, 3.472 \times 10^4, 5.556 \times 10^4$ and $3.968 \times 10^4$ N/m. Theoretical force-displacement and stiffness plots in Fig. 5d are created using these parameters ($k_1$ and $k_2$).

The experimental data show a good agreement with theoretical analysis (Fig. 5d), while there are slight discrepancy. The theoretical stiffness of the hard edge shows an insignificant decrease (no change for the pulling case) as the displacement increases; the experimental data show the increase of stiffness for the hard edge with increasing the displacement. At the low displacement range, the low stiffness is caused by the relative sliding of the fixed boundaries. But at the high displacement area, the friction, generated from the compression between hinges and triangles, dominates over the ratio of the force to displacement, leading to higher edge stiffness. The experimental stiffness of the soft edge (especially for the pushing case) is generally higher than the theoretical one since there are non-negligible surface frictions at the junctions of each unit cell.

Box plots of force-displacement are shown in Fig. S4 to illustrate surface stiffness at the soft and hard edge. Each edge is tested 15 times for both pushing and pulling it from the zero-displacement configuration with a displacement no less than 14 mm. The hard edge is around 50 times stiffer than the soft edge (see Fig. 5d), while in theory, the hard edge can get 100 times stiffer than the soft edge. This discrepancy is because that in a real system, the lattice still has friction at the contacting surfaces and bending stiffness from each triangle, resulting in the reduced difference in edge stiffness between two opposite edges. Nevertheless, these results validate that the added bistable units maintain the property of the surface stiffness of the lattice.

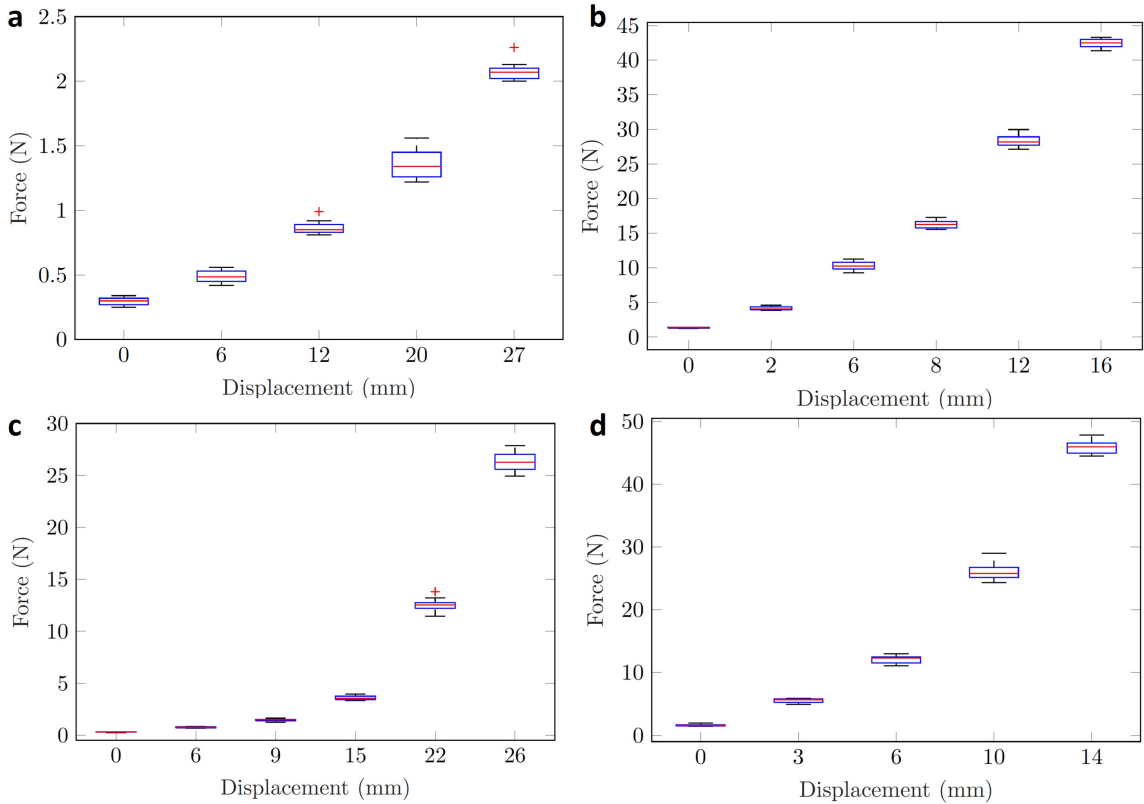

Figure S4: Box plots of force displacement relations for (a) pushing soft edge, (b) pushing hard edge, (c) pulling soft edge, and (d) pulling hard edge.

Fig. S5a and b show a 3D modeled bistable lattice design and multimaterial-3D-printed test piece, where dark colored hinges (red in (a)) are printed using Agilus30, a softer material, and light white parts are solid triangle bodies using a stiffer material, namely VeroWhite. Dark parts of the lattices are used to create hinges between the solid triangles in the lattice to resemble ideal hinged kagome lattice with zero bending stiffness. The hingeless bistable unit (3D model in Fig. S5c) 3D printed using a Nylon material (printed piece in Fig. S5d) and is used to link two solid triangles in each unit cell as a low-stiffness hinge. This hinge has two stable states, like the 3D-assembled bistable unit cell, with one state in topological polarized phase and the other one in non-polarized domain.

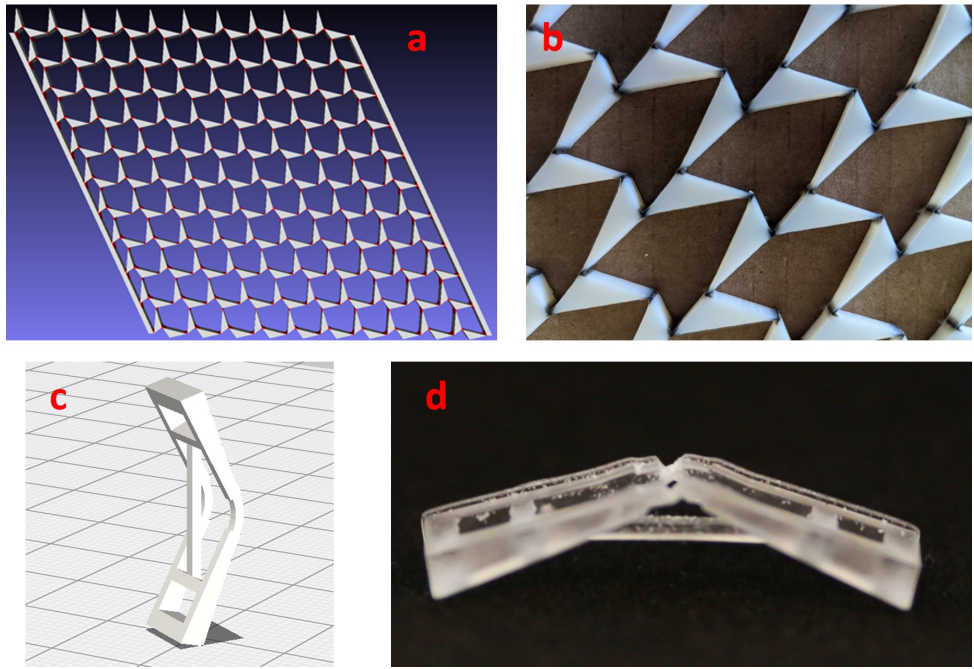

Figure S5: (a) 3-dimensional model of and (b) 3D printed multi-material bistable Maxwell lattice. (c) 3D modeled and (d) 3D printed bistable hinge.